\documentclass[aps,pre,reprint,amsmath,superscriptaddress]{revtex4-2}
\usepackage{amssymb}
\usepackage{amsmath}
\usepackage{graphicx}
\usepackage{dcolumn}
\usepackage{bm}
\usepackage[export]{adjustbox}
\usepackage{subfigure}
\usepackage{float}
\newtheorem{theorem}{Theorem}
\newtheorem{lemma}{Lemma}

\begin{document}


\title{ Rogue wave patterns\\ associated with Adler-Moser polynomials\\ in the nonlocal nonlinear Schr\"{o}dinger equation}


\author{Tingyou Wang}
\author{Zhenyun Qin}
 \email{zyqin@fudan.edu.cn}
\affiliation{School of Mathematics, Key Laboratory of Mathematics for Nonlinear Science, \\ Fudan University, Shanghai 200433, PR China}%
\author{Gui Mu}
\affiliation{School of Mathematics, Kunming University, Kunming, Yunnan 650214, PR China}
\author{Fuzhuang Zhen}
\affiliation{School of Mathematics, Key Laboratory of Mathematics for Nonlinear Science, \\ Fudan University, Shanghai 200433, PR China}%
\author{Zhijun Qiao}
\affiliation{School of Mathematical and Statistical Sciences, The University of Texas Rio Grande Valley, Edinburg, TX 78539, USA}%


\date{\today}

\begin{abstract}
In this paper, novel rogue wave patterns in the nolocal nonlinear Schr\"{o}dinger equation (NLS) are investigated by means of asymptotic analysis, including heart-pentagon, oval-trangle, and fan-trangle. It is demonstrated that when multiple free parameters get considerably large, rogue wave patterns can approximately be predicted by the root structures of Adler-Moser polynomials. These polynomials, which extend the Yablonskii-Vorob'ev polynomial hierarchy, exhibit richer geometric shapes in their root distributions. The $(x,t)$-plane is partitioned into three regions and through a combination of asymptotic results in different regions, unreported rogue wave patterns can be probed. Predicted solutions are compared with true rogue waves in light of graphical illustrations and numerical confirmation, which reveal excellent agreement between them.
\end{abstract}


\maketitle

\section{Introduction}
Rogue waves originate from oceanography and describe unexpectedly large displacements from an equilibrium position or an otherwise tranquil background \cite{khp}. In recent years, rogue waves were linked to other physical systems, including but not limited to Bose-Einstein condensates \cite{r8}, optical fibers \cite{ki,sol}, water tanks \cite{wb1,wb2,wb3}, superfluid helium \cite{sf1,sf2} and plasma \cite{pla}. Due to their physical importance, rogue waves have received intensive experimental and theoretical studies in the past decades \cite{akh,gr,pe}. Mathematically, the rational solution (also called Peregrine solution) of the focusing nonlinear Schr\"{o}dinger (NLS) equation is initially used to explain such rogue wave phenomenon in these contexts \cite{pe}. Special higher order rogue waves were obtained by Akhmediev research group using Darboux transformation \cite{aa1}, then various techniques have been taken to derive the \emph{N}th-order rational solutions of NLS equation \cite{g1,oh,du,mu2014}. Even though the mechanisms behind their generations still need to be explored, analytical expressions of rogue waves have been derived in a wide varity of integrable physical models, such as Manakov system \cite{manakov}, \textit{etc}. To construct rational solutions of integrable systems, several effective methods were successfully established, notably Darboux transformation method \cite{aa2,qin,gll}, Kadomtsev-Petviashvili (KP) hierarchy reduction technique \cite{oh,mt,hirota,yo1,mu2,wang,kg}, and inverse scattering transform \cite{ins}.

The investigation of rogue wave patterns holds significant importance, as it enables the prediction of later rogue wave events through historical rogue wave forms. Rogue wave patterns were first studied for the NLS equation, revealing that under specific parameters, its rogue waves can exhibit several specific geometric structures with symmetry and regularity, such as triangles, pentagons, and heptagons \cite{NLSPa1,NLSPa2}. These developments significantly contribute to the study of the dynamics of rogue waves and certainly make an impact on other fields as well. 

In Ref.~\cite{NLSYV}, it is remarkably noted that rogue wave patterns in the NLS equation are related to the root structures of Yablonskii-Vorob'ev polynomial hierarchies. Specifically, the patterns can approximately be predicted by the roots of Yablonskii-Vorob'ev polynomial hierarchies through rotations and contractions. Further research demonstrates that rogue waves can be associated with the root structures of more polynomial hierarchies, including Yablonskii-Vorob'ev polynomial hierarchies \cite{uniYV}, Okamoto polynomial hierarchies \cite{OM2}, generalized Wronskian Hermite polynomial hierarchies \cite{gWH2}, and Adler-Moser polynomials \cite{AM1}. For rogue waves whose $\tau$-functions have jump-3 or arbitrary jump structures,  Okamoto polynomial hierarchies and generalized Wronskian Hermite polynomial hierarchies provide effective predictive frameworks for rogue wave patterns. 

Very recently, it has been found \cite{NLNLSYV} that for the (1+1)-dimensional nonlocal NLS equation, root structures of two different Yablonskii-Vorob'ev polynomials can be contributed to predict its rogue wave patterns by dividing the $(x,t)$-plane into several regions and setting two large parameters. When multiple free parameters are appropriately scaled to large values, it indicates that the rogue wave patterns are connected to the root structures of Adler-Moser polynomials, which are more concise and diversified than the relationship with Yablonskii-Vorob'ev polynomial hierarchies \cite{AM1}. 

As an extension of the classical NLS equation, there have been extensive researches on nonlocal NLS equations \cite{NLNLS1,NLNLS2,NLNLS3,NLNLS5}. It is worthwhile to mention that the nonlocal NLS equation with $\mathcal{PT}$-symmetry proposed by Ablowitz and Musslimani
\begin{align}\label{NLNLS1}
	\mathrm{i}u_t(x,t)=u_{xx}(x,t)+2u^{2}(x,t)u^{*}(-x,t),
\end{align}
 possesses rogue wave solutions \cite{NLNLS1}, where $\mathrm{i}$ denotes the imaginary unit, $u(x,t)$ is a complex-valued function of the real variables $x$ and $t$, $*$ represents the complex conjugation, and the subscripts $x$ and $t$ indicate the derivatives with respect to $x$ and $t$, respectively. The $\mathcal{PT}$-symmetry means equation (\ref{NLNLS1}) is invariant under space inversion $x\rightarrow -x$ and time reversal $t\rightarrow -t$ \cite{PTsym}. Its rogue waves have been derived by Darboux transformation \cite{NLNLSDT}, and three families of rogue waves were reported, which means the nonlocal NLS equation exists many wider varities of rogue waves than the local one. By virtue of the KP reducion method, one of the most explicit forms for rogue wave solutions was given in Ref.~\cite{NLNLSKP}. Inspired by the above researches, we concentrate on rogue wave patterns of the nonlocal NLS equation by use of Adler-Moser polynomials to uncover more diverse geometric structures, which has not been reported in previous studies.

In this paper, it is shown that by dividing the $(x,t)$-plane into different regions and making the two sets of free parameters with large values, the shapes of rogue waves can asymptotically be predicted by root structures of Adler-Moser polynomials through a dilation and a rotation. As a consequence, some novel rogue wave patterns can be achieved in the nonlocal NLS equation.

The whole paper is organized as follows. Firstly in Section \ref{sec2}, general rogue waves of the nonlocal NLS equation are investigated with the use of the KP hierarchy reduction technique, and Adler-Moser polynomials will also be introduced. In Section \ref{sec3}, the rogue wave patterns of the nonlocal NLS equation predicted by root structures of Adler-Moser polynomials are probed, which is our main result in this paper. In Section \ref{sec4}, we center around some rogue wave patterns for specific parameters, and make comparisons between the predicted solutions and real rogue waves. In Section \ref{sec5}, we provide the analytical proof of the result emerged in Section \ref{sec3}. Last section summarizes the paper with conclusions.
\section{Preliminaries}\label{sec2}
We will focus on the nonlocal NLS equation in the form (\ref{NLNLS1}) and give its rogue wave solutions with $\tau$-functions.
\subsection{Rogue waves of the nonlocal NLS equation}
Nonlinear wave solutions in the nonlocal NLS equation such as solitons, breathers, and rogue waves has been derived by various methods before \cite{NLNLS2,NLNLS3,NLNLSDT,NLNLSKP}. The most explicit forms of the rogue wave solutions are the ones derived by the KP reduction method in Ref. \cite{NLNLSKP} and then further simplified in Ref. \cite{NLNLSYV}. 

The Schur polynomials $S_{n}(\boldsymbol{x})$ with $\boldsymbol{x}=(x_1,x_2,\cdots)$ are defined via the generating function
\begin{align}
	\sum_{k=0}^{\infty} S_k(\boldsymbol{x}) \epsilon^k=\exp \left(\sum_{k=1}^{\infty} x_k \epsilon^k\right).
\end{align} 
Specifically, we have 
\begin{align*}
	\begin{gathered}
		S_0(\boldsymbol{x})=1, \quad S_1(\boldsymbol{x})=x_1, \quad S_2(\boldsymbol{x})=\frac{1}{2} x_1^2+x_2, \ldots, \\
		S_k(\boldsymbol{x})=\sum_{l_1+2 l_2+\cdots+m l_m=k}\left(\prod_{i=1}^m \frac{x_i^{l_i}}{l_{i}!}\right),
	\end{gathered}
\end{align*}
and for $k<0$, we define $S_k(\boldsymbol{x})\equiv 0$.

\begin{lemma}\label{solution}
	The nonlocal NLS equation (\ref{NLNLS1}) has rational solutions
	\begin{align}
		u_{N_1,N_2,M_1,M_2}(x,t) = e^{-2it}\frac{\sigma_1}{\sigma_0},
	\end{align}
	with 
	\begin{align}
		\sigma_n=\left|\begin{array}{ll}
			\Gamma_{1,1}^{(n)} & \Gamma_{1,2}^{(n)} \\
			\Gamma_{2,1}^{(n)} & \Gamma_{2,2}^{(n)}
		\end{array}\right|,
	\end{align}
	and the $N_i\times M_i$ materices $\Gamma_{i,j}^{(n)}$ are given by 
	\begin{align}
		\Gamma_{i, j}^{(n)}=\left(m_{2 k-i, 2 l-j}^{(n)}|_{p=1,q=1}\right)_{1 \leq k \leq N_i, 1 \leq l \leq M_j},
	\end{align}
	where $N_1, N_2, M_1$ and $M_2$ are integers satisfing $N_1+N_2=M_1+M_2$.\\
	The definition of elements $m_{ij}^{(n)}$ are
	\begin{align}\label{mij}
		m_{i, j}^{(n)}=\sum_{v=0}^{\min (i, j)} \frac{1}{4^v} S_{i-v}\left(\boldsymbol{x}^{+}(n)+v \boldsymbol{s}\right) S_{j-v}\left(\boldsymbol{x}^{-}(n)+v \boldsymbol{s}\right)
	\end{align}
	with victors $\boldsymbol{x}^{\pm}(n)=(x_{1}^{\pm},0,x_{3}^{\pm},0,\dots)$ and $\boldsymbol{s}$ defined by
	\begin{align}
		\begin{aligned}
			x_1^{+}(n)&=x-2\mathrm{i}t+n+a_1\\ x_1^{-}(n)&=x+2\mathrm{i}t-n+b_1, \\
			x_{2 k+1}^{+}&=\frac{x-2^{2 k+1} \mathrm{i}t}{(2 k+1)!}+a_{2 k+1},\\ 
			x_{2 k+1}^{-}&=\frac{x+2^{2 k+1} \mathrm{i}t}{(2 k+1)!}+b_{2 k+1}, \\
			\sum_{k=1}^{\infty} s_k \lambda^k&=\ln \left(\frac{2}{\lambda} \tanh \frac{\lambda}{2}\right),
		\end{aligned}
	\end{align}	
	where $a_k,b_k\in \mathrm{i}\mathbb{R}$ are free parameters.
\end{lemma}

The Proof of Lemma~\ref{solution} is provided in Ref.~\cite{NLNLSYV} and Ref.~\cite{NLNLSKP}. The $\tau$ function in these rogue wave solutions possesses a $2\times2$ block matrix and the matrix element $m_{i,j}^{(n)}$ has only one summation symbol. This simplified form will facilitate the analysis of the rogue wave patterns. The number of free real parameters of the $(N_1,N_2,M_1,M_2)$-order rational solutions in Lemma~\ref{solution} are $\delta_M+\delta_N-1$, where
\begin{align}
	\begin{aligned}
		& \delta_N= \begin{cases}N_1-N_2, & \text {\quad if } N_1 \geq N_2, \\
			N_2-N_1-1, & \text {\quad if } N_1<N_2,\end{cases} \\
		& \delta_M= \begin{cases}M_1-M_2, & \text {\quad if } M_1 \geq M_2, \\
			M_2-M_1-1, & \text {\quad if } M_1<M_2.\end{cases}
	\end{aligned}
\end{align}
\subsection{The Adler-Moser polynomial and their root structures}
In this subsection, the definition of the Adler-Moser polynomials are presented, which are related to rational solutions of the Korteweg-de Vries equation and point vortex dynamics \cite{AMpoly,AMpoly2,AMpoly3}.

The Adler-Moser polynomials $\Theta_{N}(z)$ can be written as a $N\times N$ determinant
\begin{align}\label{Theta}
	\Theta_N(z)=c_N\left|\begin{array}{cccc}
		\theta_1(z) & \theta_0(z) & \cdots & \theta_{2-N}(z) \\
		\theta_3(z) & \theta_2(z) & \cdots & \theta_{4-N}(z) \\
		\vdots & \vdots & \vdots & \vdots \\
		\theta_{2 N-1}(z) & \theta_{2 N-2}(z) & \cdots & \theta_N(z)
	\end{array}\right|,
\end{align}
where $\theta_k(z)$ are Schur polynomials defined by 
\begin{align}\label{theta}
	\sum_{k=0}^{\infty} \theta_k(z) \epsilon^k=\exp \left(z\epsilon+\sum_{j=1}^{\infty} \kappa_j \epsilon^{2j+1}\right),
\end{align}
$\theta_{k}\equiv 0 $ if $k<0$, $c_N=\prod\limits^{N}_{j=1}(2j-1)!!$, and $\kappa_j\ (j\geq 1)$ are arbitrary complex constants. Note that the choice of the $\kappa_j$ constants is more flexible than the choice in Ref. \cite{AMpoly3}, which will be more convenient for our purpose.

The first few Adler-Moser polynomials are
\begin{align*}
	\begin{aligned}
		\Theta_1(z) & =z, \\
		\Theta_2(z) & =z^3-3 \kappa_1, \\
		\Theta_3(z) & =z^6-15 \kappa_1 z^3+45 \kappa_2 z-45 \kappa_1^2, \\
		\Theta_4(z) & =z^{10}-45 \kappa_1 z^7+315 \kappa_2 z^5-1575 \kappa_3 z^3, \\
		& +4725 \kappa_1 \kappa_2 z^2-4725 \kappa_1^3 z-4725 \kappa_2^2+4725 \kappa_1 \kappa_3 .
	\end{aligned}
\end{align*}
Some properties of the Adler-Moser polynomials $\Theta_{N}(z)$ will be listed below.
\begin{itemize}
	\item It is obvious from (\ref{theta}) that $\theta^{'}_k(z)=\theta_{k-1}(z)$, where the prime denotes derivative with respect to $z.$ Therefore, $\Theta_{N}(z)$ can be written as a Wronskian determinant 
	\begin{align*}
		\Theta_N(z)=\mathrm{Wr}\left[\theta_1(z), \theta_3(z), \cdots, \theta_{2 N-1}(z)\right].
	\end{align*}
	\item These Adler-Moser polynomials $\Theta_{N}(z)$ are monic with degree $\frac{N(N+1)}{2}$. It can be seen by noticing that the highest $z$ term of $\theta_{k}$ is $\frac{z^k}{k!}$, so $\Theta_{N}(z)$ in (\ref{Theta}) with element $\theta_{k}$ replaced by its highest $z$ term can be calculated as $z^{\frac{N(N+1)}{2}}$.
	\item When all constants $\kappa_j\ (j\geq 1)$ are set as $0$ except for one of them, the Alder-Moser polynoimals will reduce to the Yablonskii-Vorob'ev polynomial hierarchies, which are linked to rogue wave patterns when two free parameters are large. 
\end{itemize}

It is important to study root structures of the Adler-Moser polynomials, since they are linked to the rogue wave patterns in the later text. Their root structures are much more diverse than root structures of Yablonskii-Vorob’ev hierarchy due to the choice of the free complex parameters $\{\kappa_j|j\geq 1\}.$ Actually, when all parameters $\kappa_j$ are set as $0$ except one of them, we can figure out root structures of Yablonskii-Vorob’ev hierarchy polynomials, which have various shapes such as triangles, pentagons and heptagons. When we continuously vary the values of $\kappa_j$, the root structures undergo smooth deformation between different types. Indeed, if a root happens to be a multiple root, it can be split into several simple roots if we give a small perturbation to $\{\kappa_j|j\geq 1\}.$ Therefore, it is assumed all roots of $\Theta_{N}(z)$ are simple throughout this article, which means that $\Theta_{N}(z)$ has $\frac{N(N+1)}{2}$ simple roots.

\begin{figure}[h]
	\centering
	\includegraphics[width=3.5in]{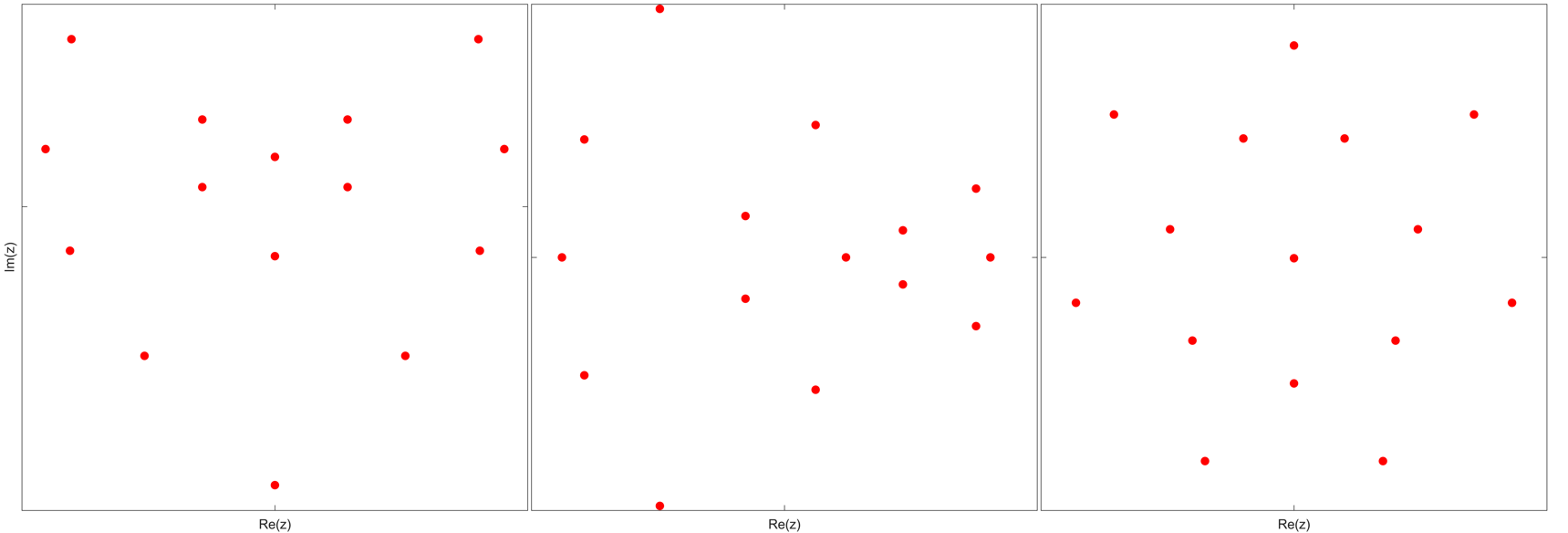}\\
	\caption{\label{AMrs}The root structures of $\Theta_{5}(z)$ for parameter values $(\kappa_1, \kappa_2, \kappa_3, \kappa_4)$ given in (\ref{examplepara}).}
\end{figure}

To demonstrate some root structures of the Adler-Moser polynomials, we select three sets of $(\kappa_1, \kappa_2, \kappa_3, \kappa_4)$ for $\Theta_{5}(z;\kappa_1, \kappa_2, \kappa_3, \kappa_4)$ as 
\begin{align}\label{examplepara}
	(\mathrm{i}, \mathrm{i}, \mathrm{i}, \mathrm{i}),\ (1,1,1,1),\ (\frac{\mathrm{i}}{4}, \frac{\mathrm{i}}{10}, -8 \mathrm{i} , \frac{\mathrm{i}}{9}).
\end{align}
Their corresponding root distributions are displayed in Fig.~\ref{AMrs}. 
\section{Analytical predictions for rogue patterns with multiple large internal parameters}\label{sec3}
In this section, the analytical predictions for rogue patterns of the nonlocal NLS equation are explored with multiple large internal parameters. Specifically, we suppose parameters $(a_3,a_5,\dots,a_{2N-1})$ and $(b_3,b_5,\dots,b_{2N-1})$ in $u_{N_1,N_2,M_1,M_2}(x,t)$ are of the following form
\begin{align}
	a_{2j+1}=\kappa_{j}A^{2j+1},\qquad b_{2j+1}=l_{j}B^{2j+1},
\end{align}
where $1\leq j\leq N-1$ and $A, B\gg 1$ are large positive constant, and $(\kappa_{1},\kappa_{2},\dots,\kappa_{N-1})$, $(l_1,l_2,\dots,l_{N-1})$ are $O(1)$ complex constants not all equal to $0$. Moreover, assume that roots of the Adler-Moser polynomials $\Theta^{[\kappa]}_N(z)$ with parameters $\{\kappa_j|j\geq 1\}$ and $\Theta^{[l]}_N(z)$ with parameters $\{l_j|j\geq 1\}$ are all simple. Then it gives rise to the following theorem about analytical predictions on the patterns of rogue wave solution $u_{N_1,N_2,M_1,M_2}(x,t)$. 
\begin{theorem}\label{Pattern}
	If all roots of Adler-Moser polynomials $\Theta^{[\kappa]}_N(z)$ and $\Theta^{[l]}_N(z)$ are simple. Under the parameter restrictions that
	\begin{align*}
		\begin{gathered}
			a_{2j+1}=\kappa_{j}A^{2j+1},\quad b_{2j+1}=l_{j}B^{2j+1},\\
			A,B\gg1\quad\text{and}\quad A^2=O(B),	
		\end{gathered}	
	\end{align*} 
	we have the following asymptotic result on the solution $u_{N_1,N_2,M_1,M_2}(x,t)$ in the $(x,t)$-plane.
	\begin{itemize}
		\item In the outer region on the $(x,t)$-plane, where $\sqrt{x^2+t^2}=O\left(B\right)$, the $(N_1,N_2,M_1,M_2)$-th order rogue waves are separated into several single rogue waves. These single rogue waves are 
		\begin{align}
			\begin{aligned}
				&u_{0110}(x-\tilde{x}_{1,0}, t-\tilde{t}_{1,0})\\
				=&e^{-2 \mathrm{i} t}\left(1-\frac{1}{\left(x-\tilde{x}_{1,0}\right)+2 \mathrm{i}\left(t-\tilde{t}_{1,0}\right)}\right),
			\end{aligned}
		\end{align}
		where their positions $(\tilde{x}_{1,0},\tilde{t}_{1,0})$ satisfy
		\begin{align*}
			\tilde{x}_{1,0}+2i \tilde{t}_{1,0}=B\tilde{z}_{1,0},
		\end{align*}
		and $\tilde{z}_{1,0}$ is a nonzero simple root of $\Theta_{\delta_M}^{[l]}(z)$. Expressed mathematically, when $\sqrt{(x-\tilde{x}_{1,0})^2+(t-\tilde{t}_{1,0})^2}=O\left(1\right)$, we have the following solution asymptotics
		\begin{align}
			\begin{aligned}
					u_{N_1,N_2,M_1,M_2}(x, t)=&u_{0110}(x-\tilde{x}_{1,0}, t-\tilde{t}_{1,0})\\&\times\left[1+O(A^{-2})\right].
			\end{aligned}
		\end{align}
		\item In the middle region on the $(x,t)$-plane, where $\sqrt{x^2+t^2}=O\left(A\right)$, the $(N_1,N_2,M_1,M_2)$-th order rogue waves are separated into several single rogue waves. These single rogue waves are 
		\begin{align}
			\begin{aligned}
				&u_{1001}(x-\tilde{x}_{2,0}, t-\tilde{t}_{2,0})\\
				=&e^{-2 \mathrm{i} t}\left(1+\frac{1}{\left(x-\tilde{x}_{2,0}\right)-2 \mathrm{i}\left(t-\tilde{t}_{2,0}\right)}\right)
			\end{aligned}
		\end{align}
		where their positions $(\tilde{x}_{2,0},\tilde{t}_{2,0})$ satisfy
		\begin{align*}
			\tilde{x}_{2,0}-2i \tilde{t}_{2,0}=A\tilde{z}_{2,0},
		\end{align*}
		and $\tilde{z}_{2,0}$ is a nonzero root of $\Theta_{\delta_N}^{[\kappa]}(z)$. Expressed mathematically, when $\sqrt{(x-\tilde{x}_{2,0})^2+(t-\tilde{t}_{2,0})^2}=O\left(1\right)$, we have the following solution asymptotics
		\begin{align}
		\begin{aligned}
				u_{N_1,N_2,M_1,M_2}(x, t)
			=&u_{1001}(x-\tilde{x}_{2,0}, t-\tilde{t}_{2,0})\\&\times\left[1+O(A^{-1})\right].
		\end{aligned}
		\end{align}
		\item In the inner region on the $(x,t)$-plane, where $\sqrt{x^2+t^2}=O\left(1\right)$. If $0$ is not a root of $\Theta_{\delta_N}^{[\kappa]}(z)$ or $\Theta_{\delta_M}^{[l]}(z)$, the $(N_1,N_2,M_1,M_2)$-th order rogue waves $u_{N_1,N_2,M_1,M_2}(x,t)$ asymptotically approach the constant background $e^{-2\mathrm{i}t}$ as $A,B\rightarrow +\infty$. Otherwise, there are three cases to be considered as follows.
		\begin{align}
			\begin{aligned}
				& u_{N_1,N_2,M_1,M_2}(x,t)\\
				= &\begin{cases}e^{-2 \mathrm{i} t}\left(1-\frac{1}{x+2 \mathrm{i}t}\right)\left[1+O(A^{-1})\right], \\ \text {\qquad\qquad if }\Theta_{\delta_N}^{[\kappa]}(0)\neq0, \Theta_{\delta_M}^{[l]}(0)=0,\\ e^{-2 \mathrm{i} t}\left(1+\frac{1}{x-2 \mathrm{i}t}\right)\left[1+O(A^{-1})\right],  \\ \text {\qquad\qquad if }\Theta_{\delta_N}^{[\kappa]}(0)=0,\Theta_{\delta_M}^{[l]}(0)\neq0,\\e^{-2 \mathrm{i} t}\left(1+\frac{16\mathrm{i}t-4}{4x^2+16t^2+1}\right)\left[1+O(A^{-1})\right], \\ \text {\qquad\qquad if }\Theta_{\delta_N}^{[\kappa]}(0)=\Theta_{\delta_M}^{[l]}(0)=0.\end{cases}
			\end{aligned}
		\end{align}
		
	\end{itemize}
\end{theorem}
\section{Comparison between true rogue patterns and analytical predictions}\label{sec4}
\subsection{Single rogue waves in Theorem~\ref{Pattern}}
\begin{figure}[h]
	\centering
	\includegraphics[width=3.5in]{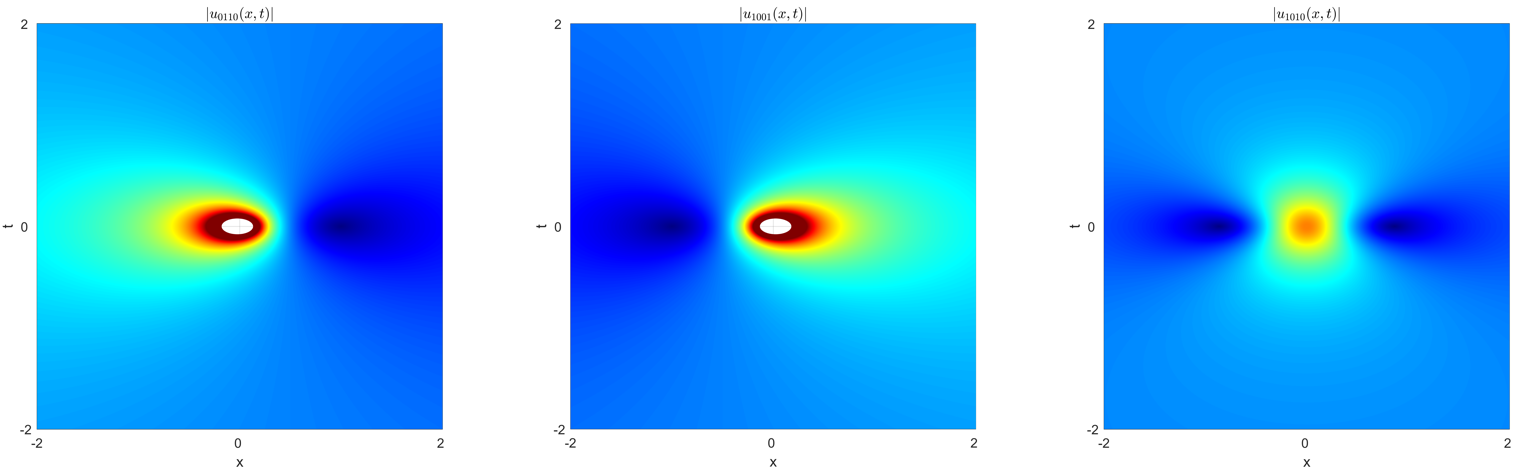}\\
	\caption{\label{single}Single rogue wave solutions in (\ref{0110})-(\ref{1010}). From left to right are $u_{0110}(x, t)$, $u_{1001}(x, t)$ , and $u_{1010}(x, t)$.}
\end{figure}

There are three types of single rogue waves mentioned in Theorem~\ref{Pattern}, and we will illustrate them in this subsection. Assume $(N_1,N_2,M_1,M_2)=(0,1,1,0),(1,0,0,1)$ and $(1,0,1,0)$ in Lemma~\ref{solution} respectively, we can figure out these specific rogue waves as 
\begin{align}
	u_{0110}(x, t)&=e^{-2 \mathrm{i} t}\left(1-\frac{1}{x+2 \mathrm{i} t}\right), \label{0110}\\
	u_{1001}(x, t)&=e^{-2 \mathrm{i} t}\left(1+\frac{1}{x-2 \mathrm{i} t}\right), \label{1001}\\
	u_{1010}(x, t)&=e^{-2 \mathrm{i} t}\left(1+\frac{16 \mathrm{i} t-4}{16 t^2+4 x^2+1}\label{1010}\right).
\end{align}
Elementary analysis reveals that $u_{1010}(x, t)$ is a Peregrine soliton while the other two are with singularities. The corresponding figures are shown in Fig.~\ref{single}

It is worth noting that only $u_{0110}(x-\tilde{x}_{1,0},t-\tilde{t}_{1,0})$ appear in the outer region and only $u_{1001}(x-\tilde{x}_{2,0},t-\tilde{t}_{2,0})$ lie in the middle region in the asymptotic expression of the rogue waves in Theorem~\ref{Pattern}. However, in the inner region $u_{0110}(x,t),u_{1001}(x,t)$ and $u_{1010}(x,t)$ are all likely to exist as asymptotic solutions.
\subsection{Rogue wave patterns in nonlocal NLS equation}
In this subsection, we will contrast the analytical predicitons with true rogue wave patterns. The analytical predictions $|u_{N_1,N_2,M_1,M_2}^{(p)}(x,t)|$ can be assembled into a simple formula with Theorem~\ref{Pattern}. For example, $\left|u_{3030}^{(p)}(x, t)\right|$ can be written as
\begin{align}\label{formula}
	\begin{aligned}
		&\left|u_{3030}^{(p)}(x, t)\right|=\hat{u}_0(x,t)\\
		+&\sum_{j=1}^{6-\delta_{\kappa}}\left(\left|u_{1001}\left(x-\hat{x}_{2,0}^{(j)}, t-\hat{t}_{2,0}^{(j)}\right)\right|-1\right)\\
		+&\sum_{j=1}^{6-\delta_{l}}\left(\left|u_{0110}\left(x-\hat{x}_{1,0}^{(j)}, t-\hat{t}_{1,0}^{(j)}\right)\right|-1\right),
	\end{aligned}
\end{align}		
where $\delta_{\kappa}=\delta_{\Theta^{[\kappa]}_3(0),0},\delta_{l}=\delta_{\Theta^{[l]}_3(0),0}$ and $\hat{u}_0(x,t)$ is the constant background $1$ or the single rogue wave in the $O(1)$ neiborhood of the origin which is determined by Theorem~\ref{Pattern}. $\left(\hat{x}_{1,0}^{(j)},\hat{t}_{1,0}^{(j)}\right)$ and $\left(\hat{x}_{2,0}^{(j)},\hat{t}_{2,0}^{(j)}\right)$ are nonzero roots of the polynomials $\Theta^{[l]}_3(z)$ and $\Theta^{[\kappa]}_3(z)$ respectively. 

\begin{figure}[h]
	\centering
	\includegraphics[width=3.5in]{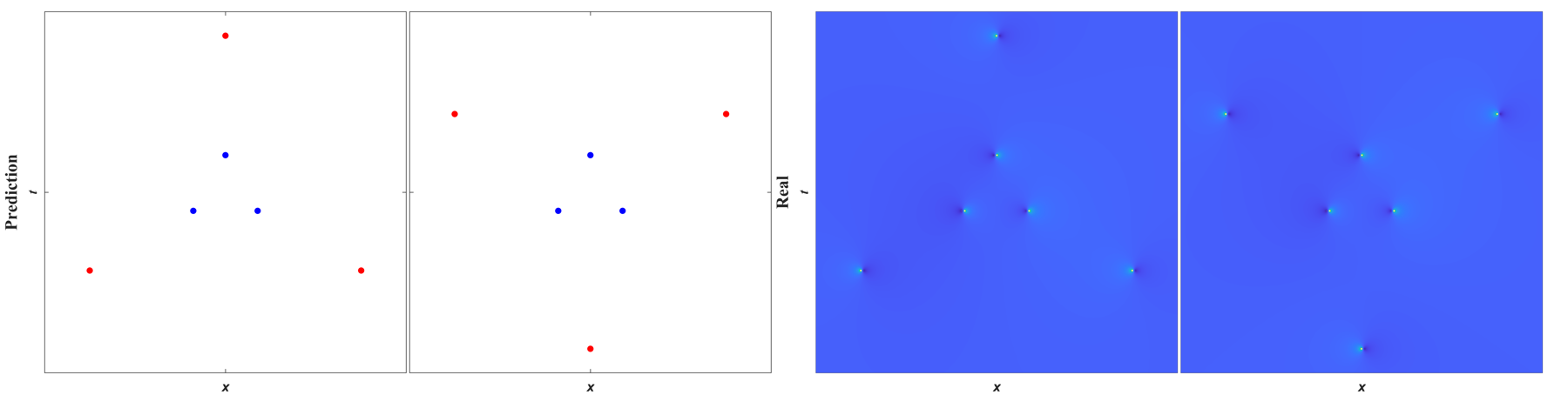}\\
	\caption{\label{2020}Comparison between analytical predictions and true $(2,0,2,0)$-th rogue solutions in nonlocal NLS equation for $A=5$ with (a) $a_3=5\cdot 5^3i,b_3=-3\cdot 25^3i$, (b) $a_3=5\cdot 5^3i,b_3=3\cdot 25^3i$ from left to right; the $(x,t)$ intervals here are $-60 \leq x \leq 60, -30 \leq t \leq 30$.}
\end{figure}

\begin{figure}[h]
	\centering
	\includegraphics[width=3.5in]{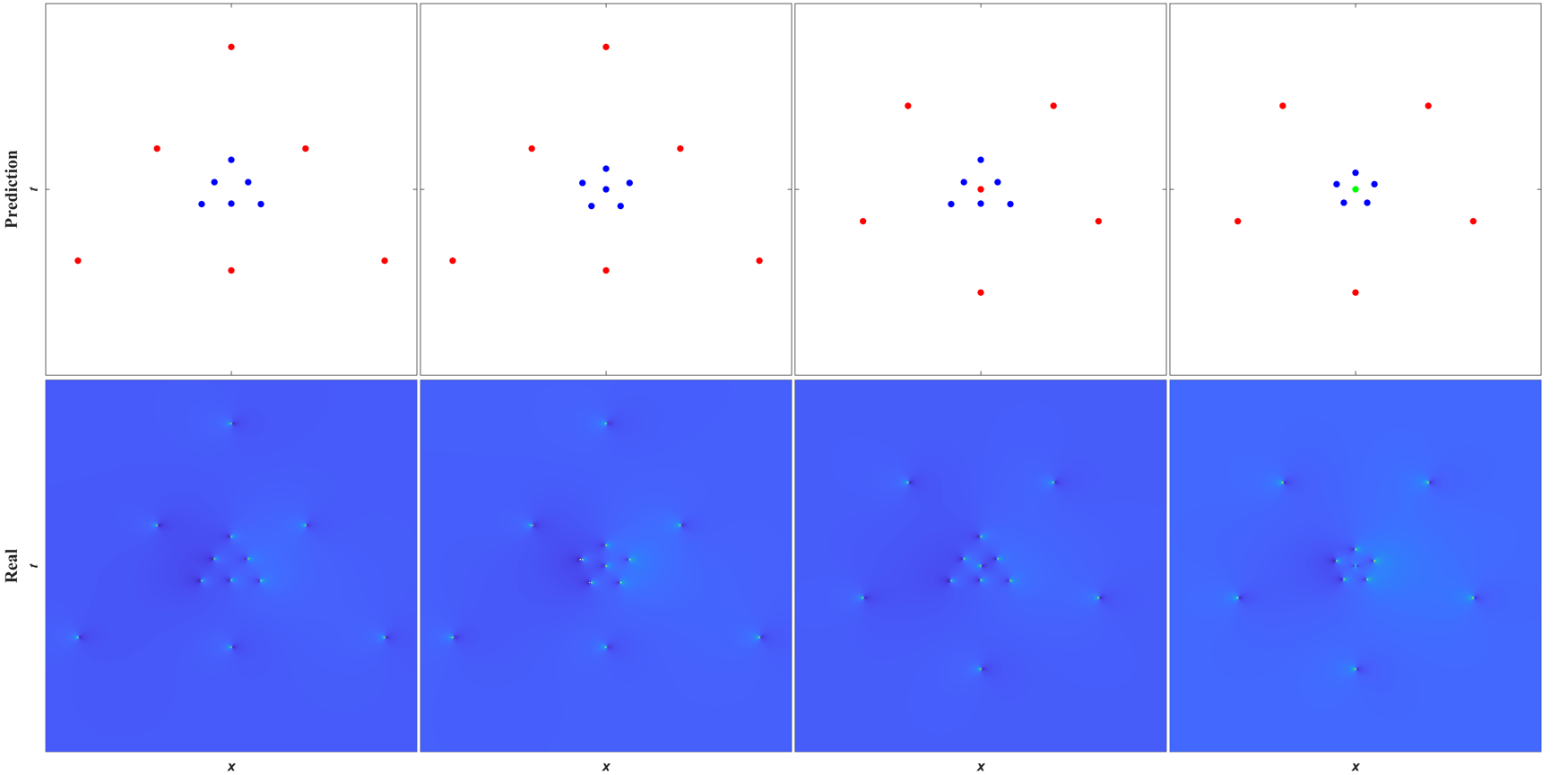}\\
	\caption{\label{3030}Comparison between analytical predictions (upper row) and true $(3,0,3,0)$-th rogue solutions (lower row) in nonlocal NLS equation for $A=5$ with (a) $a_3=5\cdot 5^3i,a_5=5^5i,b_3=-3\cdot 25^3i,b_5=25^5i$, (b) $a_3=0,a_5=3\cdot 5^5i,b_3=-3\cdot 25^3i,b_5=25^5i$, (c) $a_3=3\cdot 5^3i,a_5=5^5i,b_3=0,b_5=3\cdot 25^5i$, (d) $a_3=0,a_5=5^5i,b_3=0,b_5=3\cdot 25^5i$ from left to right; the $(x,t)$ intervals here are $-100 \leq x \leq 100, -60 \leq t \leq 60$.}
	
\end{figure}

We initiate our analysis with the case $N_1=M_1=N, N_2=M_2=0$, which means the $\tau$-function is a single block matrix. When $N = 2$ and $3$, the predicted positions of rogue wave patterns and the real ones with different free parameters $\kappa_{j}$ correspond to Fig.~\ref{2020} and Fig.~\ref{3030} respectively.

\begin{figure}[h]
	\centering
	\includegraphics[width=3.5in]{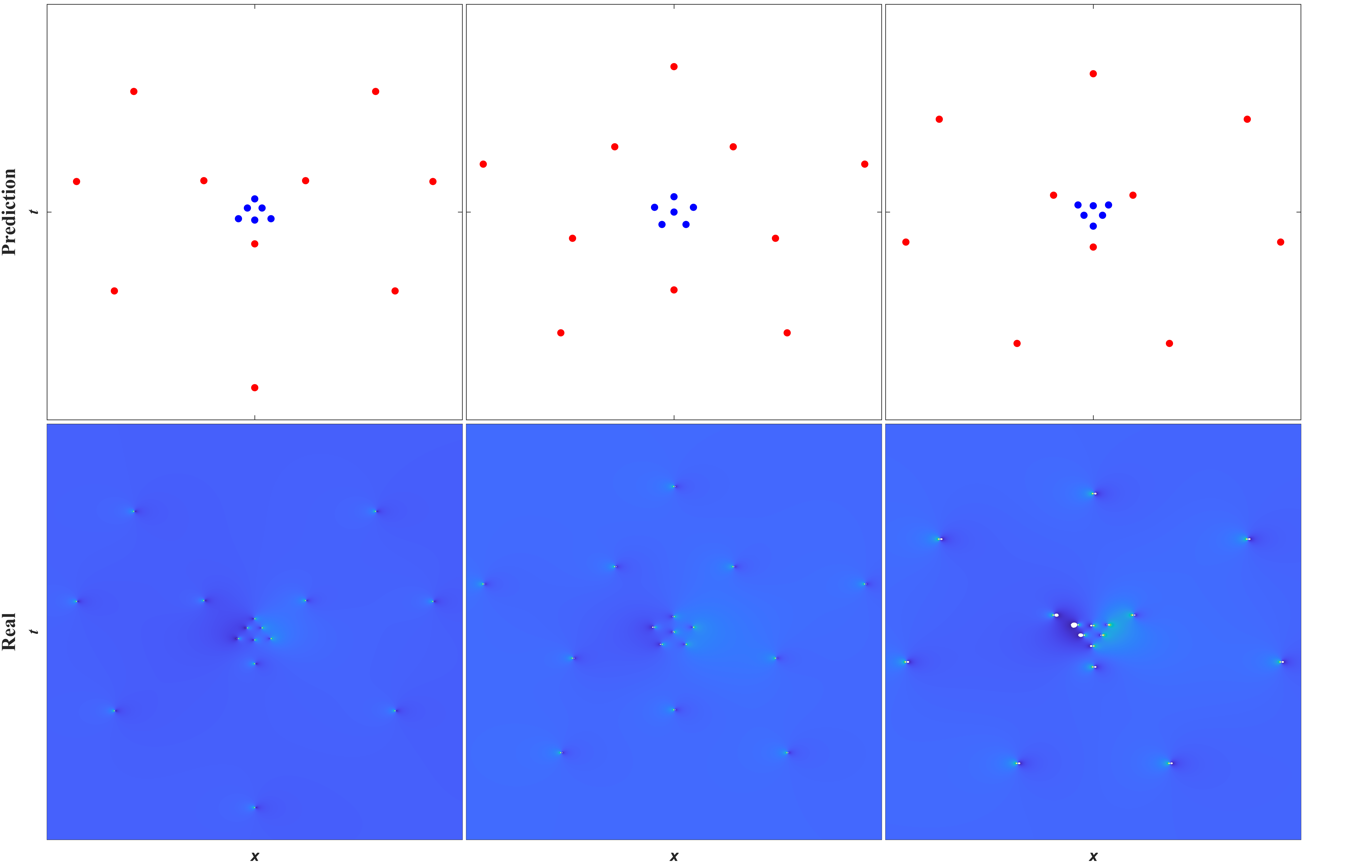}\\
	\caption{\label{0440}Comparison between analytical predictions (upper row) and true $(0,4,4,0)$-th rogue solutions (lower row) in nonlocal NLS equation for $A=6$ with (a) $a_3=-\frac{3}{5}\cdot 6^3i,a_5=\frac{1}{8}\cdot 6^5i,b_3=36^3i,b_5=2\cdot 36^5i,b_7=8\cdot 36^7i$, (b) $a_3=0,a_5=2\cdot 6^5i,b_3=\frac{1}{10}\cdot 36^3i,b_5=-3\cdot 36^5i,b_7=\frac{1}{10}\cdot 36^7i$, (c) $a_3=\frac{3}{5}\cdot 6^3i,a_5=\frac{1}{8}\cdot 6^5i,b_3=\frac{1}{4}\cdot 36^3i,b_5=\frac{3}{10}\cdot 36^5i,b_7=-8\cdot 36^7i$ from left to right; the $(x,t)$ intervals here are $-150 \leq x \leq 150, -100 \leq t \leq 100$.}
\end{figure}

\begin{figure}[h]
	\centering
	\includegraphics[width=3.5in]{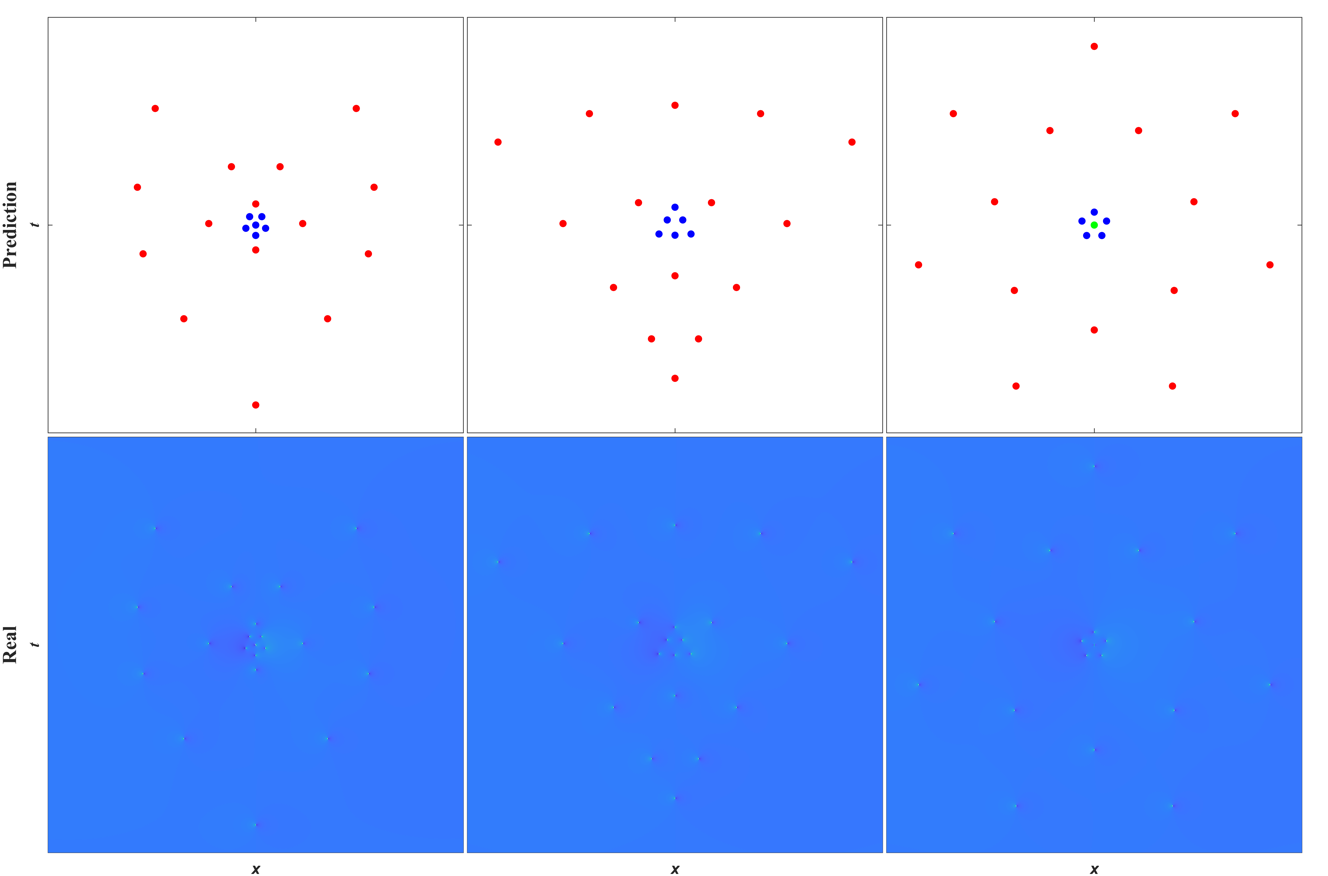}\\
	\caption{\label{4150}Comparison between analytical predictions (upper row) and true $(4,1,5,0)$-th rogue solutions (lower row) in nonlocal NLS equation for $A=7$ with (a) $a_3=0,a_5=-7^5i,b_3=\frac{3}{4}\cdot 49^3i,b_5=49^5i,b_7=49^7i,b_9=49^9i$, (b) $a_3=3\cdot 7^3,a_5=-7^5i,b_3=\frac{5}{3}\cdot 49^3i,b_5=-49^5i,b_7=\frac{5}{7}\cdot 49^7i,b_9=-\frac{5}{9}\cdot 49^9i$, (c) $a_3=0,a_5=3\cdot 7^5i,b_3=0,b_5=0,b_7=-8\cdot 49^7i,b_9=49^9i$ from left to right; the $(x,t)$ intervals here are $-300 \leq x \leq 300, -150 \leq t \leq 150$.}
\end{figure}

 In the following analysis, we investigate rogue wave patterns in general cases. As is illustrated in Fig.~\ref{0440}, for $(0,4,4,0)$-order rogue wave solutions, the selections of the free parameters $\{\kappa_{j}|j\geq1\}$ and $\{l_{j}|j\geq1\}$ in Theorem~\ref{Pattern} facilitate the generation of some novel excitation patterns, such as heart-triangle and oval-triangle. For $(4,1,5,0)$-order rogue wave solutions, there are more interesting patterns including heart-pentagon, fan-triangle as shown in Fig.~\ref{4150}. Meanwhile, patterns associated with Yablonskii-Vorob'ev polynomial hierarchies also emerge, as illustrated in the third column of Fig.~\ref{4150}.
\subsection{Numerical confirmation}
It is noted that the errors of the asymptotic prediction in the outer and middle region can be defined as 
\begin{align*}
	\text{error of single location} = \sqrt{(x_0-\tilde{x}_0)^2+(t_0-\tilde{t}_0)},
\end{align*}
with $(x_0,t_0)$ is the true position of each rogue wave and $(\tilde{x}_0,\tilde{t}_0)$ is the predicted location. Moreover, the error of the asymptotic prediction in the inner region is 
\begin{align*}
	\text{error of inner region} = \left|u_{N_1,N_2,M_1,M_2}-u^{(p)}_{N_1,N_2,M_1,M_2}\right|.
\end{align*}

\begin{figure}[h]
	\centering
	\includegraphics[width=3.5in]{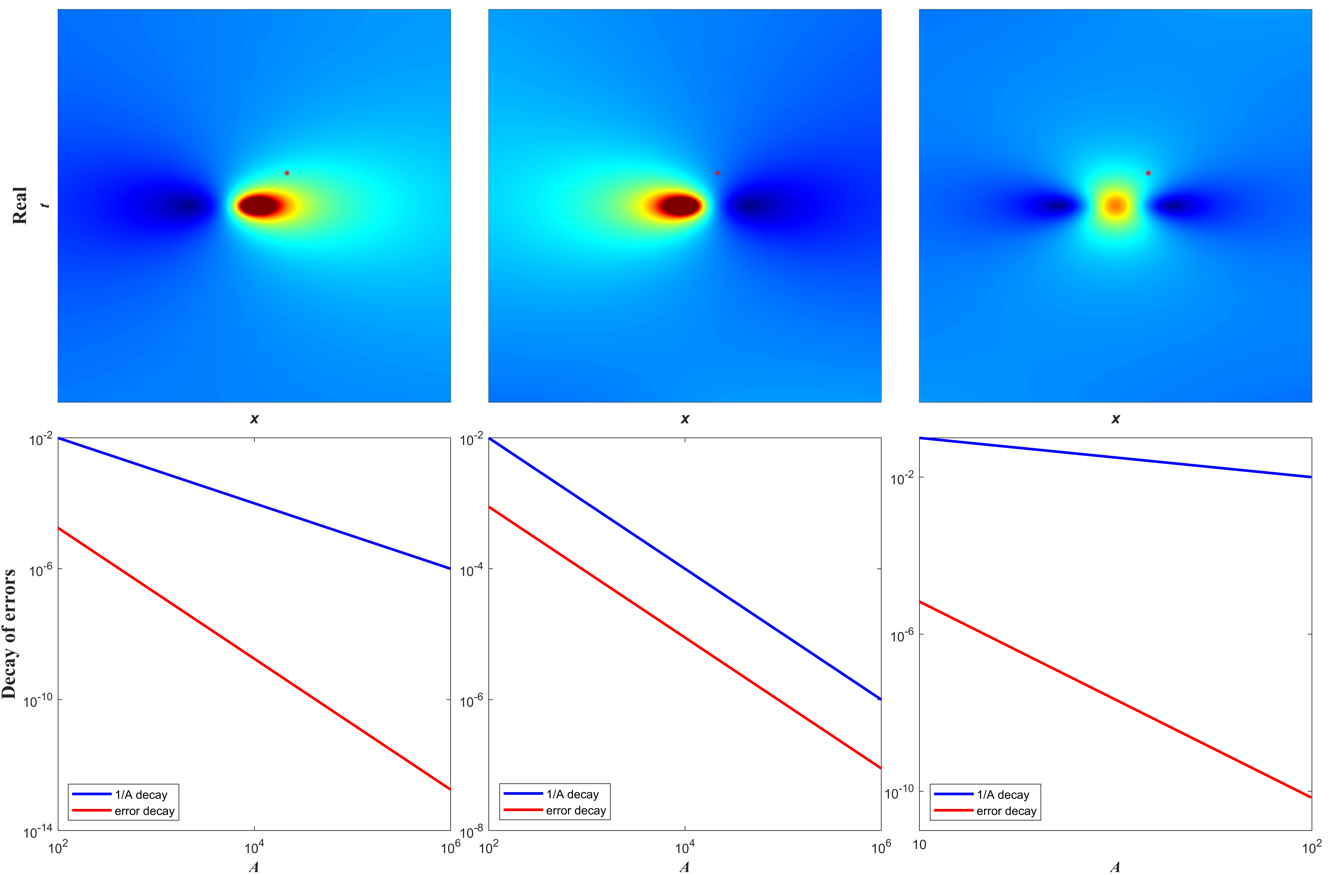}\\
	\caption{\label{error}Decay of errors in the predictions for the inner region of the $(3,0,3,0)$ -th order rogue waves in the nonlocal NLS equation at point $(\frac{1}{2},\frac{1}{2})$ as $A$ increases. The first row: real rogue waves in the inner region corresponding to the last three columns of Fig~\ref{3030}. The second row is depicted the decay of errors for the first row.}
\end{figure}

Now we numerically verify Theorem~\ref{Pattern} by comparing its predictions with real rogue waves. This comparison will be done for the $(3,0,3,0)$-order rogue waves for brevity. Quantitatively, we can measure the error of the analytical predictions versus the $A$ and $B$ value, which are shown in Fig.~\ref{error}.
\section{Proof of Theorem~\ref{Pattern}}\label{sec5}
Firstly we consider the case $N_1=M_1=N, N_2=M_2=0$, which means the $\tau$-function can be represented by a single block matrix. Then we will verify the rogue wave patterns by dividing the region into three parts. The proof for the general case where $\sigma_n$ is a $K_1\times K_2(1\leq K_1,K_2\leq 2)$ block matrix is similar to Ref. \cite{NLNLSYV}.

The main idea of the proof of Theorem~\ref{Pattern} resembles the proof of Theorem~\ref{Pattern} in Ref. \cite{NLNLSYV} for two large internal parameter case. Hence we only sketch the main differences below.

As before, we employ the determinant identities and the Laplace expansion to rewrite $\sigma_n$ as 
\begin{align}\label{Laplace}
	\begin{aligned}
		\sigma_n=& \sum_{0 \leq v_1<\cdots<v_N \leq 2 N-1} \det_{1 \leq i, j \leq N}\left[\frac{1}{2^{v_j}} S_{2 i-1-v_j}\left(\boldsymbol{x}^{+}(n)+v_j \boldsymbol{s}\right)\right] \\
		& \times \det_{1 \leq i, j \leq N}\left[\frac{1}{2^{v_j}} S_{2 i-1-v_j}\left(\boldsymbol{x}^{-}(n)+v_j \boldsymbol{s}\right)\right].
	\end{aligned}
\end{align}		
\subsection{Outer region}\label{sec5A}
When $\sqrt{x^2+t^2}=O(B)$, by recalling the definition and properties of Schur polynomials, we conclude that
\begin{align}
	\begin{aligned}
		&S_k\left(\boldsymbol{x}^{+}(n)+v_j \boldsymbol{s}\right)\\
		=&S_k\left(x_1^{+}, v s_2, x_3^{+}, v s_4, \cdots\right) \\
		=&B^{k} S_k\left(x_1^{+}B^{-1}, v s_2 B^{-2}, x_3^{+} B^{-3}, v s_4 B^{-4}, \cdots\right) \\
		=&B^{k} S_k\left((x-2\mathrm{i}t)B^{-1}, 0,0, \cdots\right)\left[1+O(B^{-1})\right] \\
		=&S_k\left(x-2\mathrm{i}t, 0,0, \cdots\right)\left[1+O(B^{-1})\right]
	\end{aligned}
\end{align}
and
\begin{align}
	\begin{aligned}
		&S_k\left(\boldsymbol{x}^{-}(n)+v_j \boldsymbol{s}\right) \\
		=&S_k\left(x_1^{-}, v s_2, x_3^{-}, v s_4, \cdots\right) \\
		=&B^{k} S_k\left(x_1^{-}B^{-1}, v s_2 B^{-2}, x_3^{-} B^{-3}, v s_4 B^{-4}, \cdots\right) \\
		=&B^{k} S_k\left((x+2\mathrm{i}t-n)B^{-1}, 0,l_1,\cdots\right)\left[1+O(B^{-2})\right] \\
		=&S_k\left(x+2\mathrm{i}t-n, 0,l_1B^3,0, \cdots\right)\left[1+O(B^{-2})\right].
	\end{aligned}
\end{align}
Then $\theta_k^{[l]}$ are related to Schur polynomials as
\begin{align}
	\begin{aligned}
		&S_k(x-2\mathrm{i}t, 0, \cdots, 0,0, \cdots)=\frac{(x-2\mathrm{i}t)^k}{k!},\\
		&S_k(x+2\mathrm{i}t-n, 0, l_1B^3,0,l_3B^5, \cdots)=B^k\theta_k^{[l]}(\tilde{z}_1),
	\end{aligned}
\end{align}
where
\begin{align}\label{tildez1}
	\tilde{z}_1=B^{-1}(x+2\mathrm{i}t-n).
\end{align}
By means of these formulas, it follows that
\begin{align}
	\begin{aligned}
		&\det_{1 \leq i, j \leq N}\left[S_{2 i-j}\left(\boldsymbol{x}^{+}(n)+v_j \boldsymbol{s}\right)\right]\\=&\gamma(x-2\mathrm{i}t)^{\frac{N(N+1)}{2}}\left[1+O(B^{-1})\right]
	\end{aligned}
\end{align}
and
\begin{align}
	\begin{aligned}
			&\det_{1 \leq i, j \leq N}\left[S_{2 i-j}\left(\boldsymbol{x}^{-}(n)+v_j \boldsymbol{s}\right)\right]\\=&c_N^{-1}B^{\frac{N(N+1)}{2}}\Theta_{N}^{[l]}(\tilde{z}_1)\left[1+O(B^{-2})\right],
	\end{aligned}
\end{align}
with $\gamma$ is a real constant and $c_N=\prod\limits_{j=1}^{N}(2j-1)!!$.\\
Since the highest order term of $\sigma_n$ in (\ref{Laplace}) comes from the index choices of $v_j=j-1$, it yields
\begin{align}\label{outersigma}
	\begin{aligned}
		\sigma_n(x, t) =& \frac{\gamma c_N^{-1}}{2^{N(N-1)}} B^{\frac{N(N+1)}{2}}(x-2\mathrm{i}t)^{\frac{N(N+1)}{2}}\\ &\times\Theta_N^{\left[l\right]}\left(\tilde{z}_1\right)\left[1+O(B^{-1})\right].
	\end{aligned}
\end{align}
It is obvious to see that in the outer region 
\begin{align*}
	\frac{\sigma_1}{\sigma_0}\sim 1
\end{align*}
except at or near $(x,t)$ locations $(\tilde{x}_{1,0},\tilde{t}_{1,0})$, where
\begin{align}
	\tilde{z}_{1,0}=B^{-1}\left(\tilde{x}_{1,0}+2\mathrm{i}\tilde{t}_{1,0}\right)	
\end{align}
is a nonzero root of the polynomial $\Theta_N^{[l]}(\tilde{z}_1)$.

When $(x,t)$ is in the $O(1)$ neighborhood of $(\tilde{x}_{1,0},\tilde{t}_{1,0})$, we expand $\Theta_{N}^{[l]}(\tilde{z}_1)$ around $\tilde{z}_1=\tilde{z}_{1,0}$. Recalling $\Theta_{N}^{[l]}(\tilde{z}_{1,0})=0$ implies
\begin{align}\label{expansion1}
	\begin{aligned}
		\Theta_{N}^{[l]}(\tilde{z}_1)=&B^{-1}\left[(x-\tilde{x}_{1,0})+2\mathrm{i}(t-\tilde{t}_{1,0})-n\right]
	\\&\times\left(\Theta_{N}^{[l]}\right)^{'}(\tilde{z}_{1,0})\left[1+O(B^{-1})\right].
	\end{aligned}
\end{align}
Inserting this equation into (\ref{outersigma}), it follows that
\begin{align}\label{outersigmafinal}
	\begin{aligned}
		\sigma_n(x, t) =& \frac{\gamma c_N^{-1}}{2^{N(N-1)}} B^{\frac{N(N+1)-2}{2}}\\
		&\times\left[(x-\tilde{x}_{1,0})+2\mathrm{i}(t-\tilde{t}_{1,0})-n\right]\\
		&\times(x-2\mathrm{i}t)^{\frac{N(N+1)}{2}} \left(\Theta_{N}^{[l]}\right)^{'}(\tilde{z}_{1,0})\\&\times\left[1+O(B^{-1})\right].
	\end{aligned}
\end{align}
It should be noted that the highest order term in (\ref{Laplace}) of the index choices $v=(1,2,\cdots,N-2,N)$ is not the leading order term in this case, which is different from the local cases. Since the root $\tilde{z}_{1,0}$ has been assumed simple, $\left(\Theta_{N}^{[l]}\right)^{'}(\tilde{z}_{1,0})\neq 0$. Thus, the above leading order term asymptotics for $\sigma_n$ does not vanish. Therefore, when $A,B$ is large and $(x,t)$ in the $O(1)$ neighborhood of $(\tilde{x}_{1,0},\tilde{t}_{1,0})$, by setting $n$ to $0$ and $1$ respectively in (\ref{outersigmafinal}), we arrive at
\begin{align}
	\begin{aligned}
		&u_{N_1,N_2,M_1,M_2}(x,t)\\
		=&e^{-2\mathrm{i}t}\left(1-\frac{1}{\left(x-\tilde{x}_{1,0}\right)+2 \mathrm{i}\left(t-\tilde{t}_{1,0}\right)}\right)\\&\times\left[1+O(B^{-1})\right],	
	\end{aligned}
\end{align}
which is the single rogue wave $u_{0110}(x-\tilde{x}_{1,0},t-\tilde{t}_{1,0})$, and the error of this prediction is $O(B^{-1})$.
\subsection{Middle region}\label{sec5B}
In the middle region $\sqrt{x^2+t^2}=O(A)$, some additional restrictions on the large parameters $A$ and $B$ should be imposed to avoid confusion between the root systems of $\Theta_N^{[\kappa]}(z)$ and $\Theta_{N}^{[l]}(z)$.\\
For a fixed integer $N$, it is denoted that
\begin{align}
	R_{N}=\min \left\{\left|\tilde{z}_{1}\right| \mid \tilde{z}_{1} \in \mathbb{C} \backslash\{0\}, \Theta_N^{\left[l\right]}\left(\tilde{z}_{1}\right)=0\right\},
\end{align} 
and choose sufficiently large $A$ and $B$ such that
\begin{align*}
	O(A)\ll BR_N=O(A^2)=O(B).
\end{align*}
This indicates that $\Theta_{N}^{[l]}(\tilde{z}_{1})$ has no zeros in the middle region $\sqrt{x^2+t^2}=O(A)$ in this case. Similar to the case in the outer region, it leads to
\begin{align}
	\begin{aligned}
		&S_k\left(\boldsymbol{x}^{+}(n)+v_j \boldsymbol{s}\right) \\=&S_k\left(x_1^{+}, v s_2, x_3^{+}, v s_4, \cdots\right) \\
		=&A^{k} S_k\left(x_1^{+}A^{-1}, v s_2 A^{-2}, x_3^{+} A^{-3},\cdots\right) \\
		=&A^{k} S_k\left((x-2\mathrm{i}t+n)A^{-1}, 0,\kappa_{1}, 0, \cdots\right)\left[1+O(A^{-2})\right] \\
		=&S_k\left(x-2\mathrm{i}t+n, 0,\kappa_1A^3, 0, \cdots\right)\left[1+O(A^{-2})\right],
	\end{aligned}
\end{align}
and 
\begin{align}
	\begin{aligned}
			S_k\left(\boldsymbol{x}^{-}(n)+v_j \boldsymbol{s}\right) =&S_k\left(x+2\mathrm{i}t-n, 0,l_1B^3, 0, \cdots\right)\\&\times\left[1+O(B^{-2})\right].
	\end{aligned}
\end{align}
Therefore, these Schur polynomials are related to $\theta_k^{[\kappa]}(z)$ and $\theta_k^{[l]}(z)$ as 
\begin{align}\label{asy_middle}
	\begin{aligned}
		&S_k(x-2\mathrm{i}t+n, 0,\kappa_1A^3,0,\kappa_3A^5, \cdots)=A^k\theta_k^{[\kappa]}(\tilde{z}_2),\\
		&S_k(x+2\mathrm{i}t-n, 0, l_1B^3,0,l_3B^5, \cdots)=B^k\theta_k^{[l]}(\tilde{z}_1),
	\end{aligned}
\end{align}
where 
\begin{align}
	\tilde{z}_2=A^{-1}(x-2\mathrm{i}t+n),
\end{align}
and $\tilde{z}_1$ has been given by (\ref{tildez1}).\\
With the help of these formulas, it can be deduced that
\begin{align}
	\begin{aligned}
		&\det_{1 \leq i, j \leq N}\left[S_{2 i-j}\left(\boldsymbol{x}^{+}(n)+v_j \boldsymbol{s}\right)\right]\\=&c_N^{-1}A^{\frac{N(N+1)}{2}}\Theta_{N}^{[\kappa]}(\tilde{z}_2)\left[1+O(A^{-2})\right]
	\end{aligned}
\end{align}
and
\begin{align}
	\begin{aligned}
		&\det_{1 \leq i, j \leq N}\left[S_{2 i-j}\left(\boldsymbol{x}^{-}(n)+v_j \boldsymbol{s}\right)\right]\\=&c_N^{-1}B^{\frac{N(N+1)}{2}}\Theta_{N}^{[l]}(\tilde{z}_1)\left[1+O(A^{-4})\right].
	\end{aligned}
\end{align}
The highest order term of $\sigma_n$ in this case still comes from the index choices of $v_j=j-1$, it is concluded that
\begin{align}\label{middlesigma}
	\begin{aligned}
		\sigma_n(x, t) =& \frac{c_N^{-2}}{2^{N(N-1)}} A^{\frac{N(N+1)}{2}}B^{\frac{N(N+1)}{2}}\\&\times\Theta_N^{\left[\kappa\right]}\left(\tilde{z}_2\right)\Theta_N^{\left[l\right]}\left(\tilde{z}_1\right)\left[1+O(A^{-2})\right].
	\end{aligned}
\end{align}
This immediately implies
\begin{align*}
	\frac{\sigma_1}{\sigma_0}\sim 1
\end{align*}
except at or near $(x,t)$ locations $(\tilde{x}_{2,0},\tilde{t}_{2,0})$, where
\begin{align}
	\tilde{z}_{2,0}=A^{-1}\left(\tilde{x}_{2,0}-2\mathrm{i}\tilde{t}_{2,0}\right)	
\end{align}
is a nonzero root of the polynomial $\Theta_N^{[\kappa]}(\tilde{z}_2)$.

When $(x,t)$ is in the $O(1)$ neighborhood of $(\tilde{x}_{2,0},\tilde{t}_{2,0})$, a similar expansion as the outer region can be fulfilled as
\begin{align}\label{expansion2}
	\begin{aligned}
		\Theta_{N}^{[\kappa]}(\tilde{z}_2)=&A^{-1}\left[(x-\tilde{x}_{2,0})-2\mathrm{i}(t-\tilde{t}_{2,0})+n\right]\\
		&\times\left(\Theta_{N}^{[\kappa]}\right)^{'}(\tilde{z}_{2,0})\left[1+O(A^{-1})\right].
	\end{aligned}
\end{align}
Inserting it into (\ref{middlesigma}), this yields
\begin{align}\label{middlesigmafinal}
	\begin{aligned}
		\sigma_n(x, t) = &\frac{c_N^{-2}}{2^{N(N-1)}} A^{\frac{N(N+1)-2}{2}}B^{\frac{N(N+1)}{2}}\\&\times\left[(x-\tilde{x}_{1,0})-2\mathrm{i}(t-\tilde{t}_{1,0})+n\right]\\
		&\times\left(\Theta_{N}^{[\kappa]}\right)^{'}(\tilde{z}_{2,0})\Theta_{N}^{[l]}(\tilde{z}_{2,0})\left[1+O(A^{-1})\right].
	\end{aligned}
\end{align}
Since the root $\tilde{z}_{2,0}$ has been supposed simple and $\Theta_{N}^{[\l]}(\tilde{z}_{1})$ has no zeros in the middle region, $\left(\Theta_{N}^{[l]}\right)^{'}(\tilde{z}_{2,0})\Theta_{N}^{[l]}(\tilde{z}_{2,0})\neq 0$. Thus, the above leading order term asymptotics for $\sigma_n$ does not vanish. Therefore, when $A,B$ is large and $(x,t)$ in the $O(1)$ neighborhood of $(\tilde{x}_{2,0},\tilde{t}_{2,0})$, (\ref{middlesigmafinal}) implies
\begin{align}
	\begin{aligned}
		&u_{N_1,N_2,M_1,M_2}(x,t) \\
		=&e^{-2\mathrm{i}t}\left(1+\frac{1}{\left(x-\tilde{x}_{2,0}\right)-2 \mathrm{i}\left(t-\tilde{t}_{2,0}\right)}\right)\\&\times\left[1+O(A^{-1})\right],
	\end{aligned}
\end{align}
which is the single rogue wave $u_{1001}(x-\tilde{x}_{2,0},t-\tilde{t}_{2,0})$, and the error of this prediction is $O(A^{-1})$.
\subsection{Inner Region}\label{sec5C}
In the inner region $\sqrt{x^2+t^2}=O(1)$, there are several cases to be considered for analytical predction on the rogue wave patterns.\\
If $0$ is neither a root of $\Theta_{N}^{[\kappa]}(z)$ nor a root of $\Theta_{N}^{[l]}(z)$, it gives rise to 
\begin{align}
	\begin{aligned}
		\sigma_n(x, t) =& \frac{c_N^{-2}}{2^{N(N-1)}} A^{\frac{N(N+1)}{2}}B^{\frac{N(N+1)}{2}}\\&\times\Theta_N^{\left[\kappa\right]}\left(0\right)\Theta_N^{\left[l\right]}\left(0\right)\left[1+O(A^{-1})\right].
	\end{aligned}
\end{align}
In this case, the highest order term asymptotic for $\sigma_n$ does not vanish. Thus it results in
\begin{align}
	u_{N_1,N_2,M_1,M_2}(x,t) = e^{-2\mathrm{i}t}\frac{\sigma_1}{\sigma_0}\sim e^{-2\mathrm{i}t}
\end{align} 
as $A,B \rightarrow +\infty$.

If $0$ is a root of $\Theta_{N}^{[\kappa]}(z)$ but not a root of $\Theta_{N}^{[l]}(z)$, through a similar
calculation in the middle region, it yields
\begin{align}\label{innersigma1}
\begin{aligned}
		\sigma_n(x, t) =& \frac{c_N^{-2}}{2^{N(N-1)}} A^{\frac{N(N+1)}{2}}B^{\frac{N(N+1)}{2}}\\&\times\Theta_N^{\left[\kappa\right]}\left(\tilde{z}_2\right)\Theta_N^{\left[l\right]}\left(0\right)\left[1+O(A^{-1})\right].
\end{aligned}
\end{align}
Inserting the expansion
\begin{align}\label{expansioninner1}
	\begin{aligned}
		\Theta_{N}^{[\kappa]}(\tilde{z}_2)=&A^{-1}\left(x-2\mathrm{i}t+n\right)\\&\times\left(\Theta_{N}^{[\kappa]}\right)^{'}(0)\left[1+O(A^{-1})\right]
	\end{aligned}
\end{align}
into (\ref{innersigma1}), it follows
\begin{align}
	\begin{aligned}
		\sigma_n(x, t) =&\frac{c_N^{-2}}{2^{N(N-1)}} A^{\frac{N(N+1)-2}{2}}B^{\frac{N(N+1)}{2}}\left(x-2\mathrm{i}t+n\right)\\
		&\times\left(\Theta_{N}^{[\kappa]}\right)^{'}(0)\Theta_{N}^{[l]}(0)\left[1+O(A^{-1})\right].
	\end{aligned}
\end{align}
Therefore, the analytical predction of rogue wave solution in this case is
\begin{align}
	\begin{aligned}
		u_{N_1,N_2,M_1,M_2}(x,t)=&e^{-2 \mathrm{i} t}\left(1+\frac{1}{x-2 \mathrm{i}t}\right)\left[1+O(A^{-1})\right].
	\end{aligned}
\end{align}

If $0$ is a root of $\Theta_{N}^{[l]}(z)$ but not a root of $\Theta_{N}^{[\kappa]}(z)$, $\sigma_n(x,t)$ satisfies
\begin{align}\label{innersigma2}
	\begin{aligned}
		\sigma_n(x, t) =& \frac{c_N^{-2}}{2^{N(N-1)}} A^{\frac{N(N+1)}{2}}B^{\frac{N(N+1)}{2}}\\&\times\Theta_N^{\left[\kappa\right]}\left(0\right)\Theta_N^{\left[l\right]}\left(\tilde{z}_1\right)\left[1+O(A^{-1})\right].
	\end{aligned}
\end{align}
We can insert the expansion
\begin{align}\label{expansioninner2}
	\begin{aligned}
		\Theta_{N}^{[l]}(\tilde{z}_1)=&B^{-1}\left(x+2\mathrm{i}t-n\right)\\&\times\left(\Theta_{N}^{[l]}\right)^{'}(0)\left[1+O(A^{-2})\right]
	\end{aligned}
\end{align}
into (\ref{middlesigma}) to get 
\begin{align}
	\begin{aligned}
		\sigma_n(x, t) =& \frac{c_N^{-2}}{2^{N(N-1)}} A^{\frac{N(N+1)}{2}}B^{\frac{N(N+1)-2}{2}}\\&\times\left(x+2\mathrm{i}t-n\right)\left(\Theta_{N}^{[l]}\right)^{'}(0)\Theta_{N}^{[\kappa]}(0)\\
		& \times\left[1+O(A^{-1})\right].
	\end{aligned}
\end{align}
The analytical predction of rogue wave solution in this case is
\begin{align}
\begin{aligned}
		u_{N_1,N_2,M_1,M_2}(x,t)=&e^{-2 \mathrm{i} t}\left(1-\frac{1}{x+2 \mathrm{i}t}\right)\left[1+O(A^{-1})\right].
\end{aligned}
\end{align}

In the last case that $0$ is both a root of $\Theta_{N}^{[\kappa]}(z)$ and a root of $\Theta_{N}^{[l]}(z)$, the leading order term asymptotic for $\sigma_n$ is different from the previous cases. The dominant contribution in the Laplace expansion (\ref{Laplace}) of $\sigma_n$ comes from two index choices  $v=(0,1,\cdots,N)$ and $v=(0,1,\cdots,N-2,N)$.

With the first index choice, the determinant involving $\boldsymbol{x}^{+}(n)$ inside the summation of (\ref{Laplace}) is asymptotically
\begin{align}\label{asyx+}
	\frac{c_N^{-1}}{2^{\frac{N(N-1)}{2}}} A^{\frac{N(N+1)}{2}}\Theta_N^{\left[\kappa\right]}\left(\tilde{z}_{2}\right)\left[1+O(A^{-2})\right].
\end{align}
Inserting the expansion (\ref{expansioninner1}) into (\ref{asyx+}), the determinant involving $\boldsymbol{x}^{+}(n)$ becomes
\begin{align}
	\begin{aligned}
		\frac{c_N^{-1}}{2^{\frac{N(N-1)}{2}}} A^{\frac{N(N+1)-2}{2}}&\left(x-2it+n\right)\left(\Theta_{N}^{[\kappa]}\right)^{'}(0)\left[1+O(A^{-1})\right].
	\end{aligned}
\end{align}
Similarly, the determinant involving $\boldsymbol{x}^{-}(n)$ inside this summation can be written as
\begin{align}
	\frac{c_N^{-1}}{2^{\frac{N(N-1)}{2}}} B^{\frac{N(N+1)-2}{2}}\left(x+2it-n\right)\left(\Theta_{N}^{[l]}\right)^{'}(0)\left[1+O(A^{-2})\right].
\end{align}

Next, we consider the contribution from the second index choice of $v=(0,1,\cdots,N-2,N)$. For this choice, the determinant involving $\boldsymbol{x}^{+}(n)$ becomes
\begin{align}
	\frac{1}{2}\frac{c_N^{-1}}{2^{\frac{N(N-1)}{2}}} A^{\frac{N(N+1)-2}{2}}\left(\Theta_{N}^{[\kappa]}\right)^{'}(0)\left[1+O(A^{-1})\right],
\end{align}
and the determinant involving $\boldsymbol{x}^{-}(n)$ inside this summation becomes
\begin{align}
	\frac{1}{2}\frac{c_N^{-1}}{2^{\frac{N(N-1)}{2}}} B^{\frac{N(N+1)-2}{2}}\left(\Theta_{N}^{[l]}\right)^{'}(0)\left[1+O(A^{-2})\right].
\end{align}
Summarizing the above two dominant contributions in the Laplace expansion (\ref{Laplace}), we arrive at
\begin{align}
	\begin{aligned}
		\sigma_n(x, t) =& \frac{c_N^{-2}}{2^{N(N-1)}} A^{\frac{N(N+1)-2}{2}}B^{\frac{N(N+1)-2}{2}}\\&\times\left(x^2+4t^2+4\mathrm{i}nt-n^2+\frac{1}{4}\right)\\
		& \times\left(\Theta_{N}^{[\kappa]}\right)^{'}(0)\left(\Theta_{N}^{[l]}\right)^{'}(0)\left[1+O(A^{-1})\right].
	\end{aligned}
\end{align}
Since the root $0$ has been assumed simple, $\left(\Theta_{N}^{[\kappa]}\right)^{'}(0)\left(\Theta_{N}^{[l]}\right)^{'}(0)\neq0$. Thus, the above leading-order term asymptotics for $\sigma_n$ does not vanish. Therefore, it leads to 
\begin{align}
	\begin{aligned}
		u_{N_1,N_2,M_1,M_2}(x,t)=&e^{-2 \mathrm{i} t}\left(1+\frac{16\mathrm{i}t-4}{4x^2+16t^2+1}\right)\\&\times\left[1+O(A^{-1})\right],
	\end{aligned}
\end{align}
which is the single rogue wave $u_{1010}(x,t)$ and the error of this predction is $O(A^{-1})$.

Finally, we will give a brief proof for the more general case where $\sigma_n$ is a $K_1\times K_2(1\leq K_1,K_2\leq 2)$ block matrix. As mentioned in Ref.~\cite{NLNLSYV}, the $(N_1,N_2,M_1,M_2)$-th order rational solution $u_{N_1,N_2,M_1,M_2}(x,t)$ is equivalent to a $(\delta_N,0,\hat{M_1},\hat{M_2})$-th order rational solution when $\delta_N\geq \delta_N$ and to a $(\hat{N_1},\hat{N_2},\delta_M,0)$-th order rational solution when $\delta_N<\delta_M$. Without loss of generality, we only consider the case where $\delta_N<\delta_M$. In this case, actually we just need to focus on the rogue wave pattern of $u_{\hat{N_1},\hat{N_2},\delta_M,0}(x,t)$.

In the middle region $\sqrt{x^2+t^2}=O(A)$, we can still make a similar approximation as before. Specifically, $\sigma_n$ of the $(\hat{N_1},\hat{N_2},\delta_M,0)$-th order rogue wave solution can be expressed as
\begin{align}
	\sigma_n=\left|\begin{array}{l}
		\mathbf{m}_{1,1}^{(n)} \\
		\mathbf{m}_{2,1}^{(n)}
	\end{array}\right|=\left|\begin{array}{cc}
		\mathbf{O}_{\hat{N}_1 \times \hat{N}} & \Phi_{1, \hat{N}_1 \times 2 \hat{N}} \\
		\mathbf{O}_{\hat{N}_2 \times \hat{N}} & \Phi_{2, \hat{N}_2 \times 2 \hat{N}} \\
		-\Psi_{2 \hat{N} \times \hat{N}} & \mathbf{I}_{2 \hat{N} \times 2 \hat{N}}
	\end{array}\right|,
\end{align}
with $\mathbf{m}_{i,j}^{(n)}$ are given in (\ref{mij}), $\hat{N}=\hat{N_1}+\hat{N_2}=\delta_M$, $\Psi_{i,j}=2^{-(i-1)} S_{2 j-i}\left[\boldsymbol{x}^{-}(n)+(i-1) \boldsymbol{s}\right]$ and
\begin{align}
	\Phi_{k,i,j}=\left(\frac{1}{2^{j-1}} S_{2 i-j-k+1}\left[\boldsymbol{x}^{+}(n)+(j-1) \boldsymbol{s}\right]\right).
\end{align}
By aid of the Laplace expansion, we can rewrite $\sigma_n$ as
\begin{align}\label{Laplace2}
	\begin{aligned}
		\sigma_n=& \sum_{0 \leq v_1<\cdots<v_{\hat{N}} \leq 2 \hat{N}-1}\det_{1 \leq i, j \leq \hat{N}}\left[\frac{1}{2^{v_j}} S_{2 i-1-v_j}\left(\boldsymbol{x}^{-}(n)+v_j \boldsymbol{s}\right)\right].\\
		& \times \det\left\{\begin{array}{l}
			\left[\frac{1}{2^{v_j}} S_{2 i-1-v_j}\left(\boldsymbol{x}^{+}(n)+v_j \boldsymbol{s}\right)\right]_{1\leq i\leq \hat{N_1},1\leq j\leq \hat{N}} \\
			\left[\frac{1}{2^{v_j}} S_{2 i-2-v_j}\left(\boldsymbol{x}^{+}(n)+v_j \boldsymbol{s}\right)\right]_{1\leq i\leq \hat{N_2},1\leq j\leq \hat{N}}
		\end{array}\right\}.
	\end{aligned}
\end{align}	
By using (\ref{asy_middle}), it follows that the highest order term of $\sigma_n$ in this case still comes from the index choices of $v_j=j-1$. When $A,B\gg 1$, it can be derived that
\begin{align}\label{middlesigma2}
	\begin{aligned}
		\sigma_n(x, t) =& \frac{c_{\hat{N}}^{-1}}{2^{\hat{N}(\hat{N}-1)}} B^{\frac{\hat{N}(\hat{N}+1)}{2}}\Theta_{\hat{N}}{\left[l\right]}\left(\tilde{z}_1\right)\\&\times P_{\hat{N}}^{\left[\kappa\right]}\left(\tilde{z}_2\right)\left[1+O(A^{-2})\right],
	\end{aligned}
\end{align}
where
\begin{align}\label{generaldeterminant}
	P_{\hat{N}}^{\left[\kappa\right]}\left(\tilde{z}_2\right)=\left|\begin{array}{l}
		\hat{P}_{1, \hat{N}_1 \times \hat{N}}^{\left[\kappa\right]} \\
		\hat{P}_{2, \hat{N}_2 \times \hat{N}}^{\left[\kappa\right]}
	\end{array}\right|,
\end{align}
with
\begin{align}
	\hat{P}_k^{\left[\kappa\right]}=\left(A^{2i-j-k+1} \theta_{2 i-j-k+1}^{\left[\kappa\right]}\left(\tilde{z_2}\right)\right)_{1 \leq i \leq \hat{N}_k, l \leq j \leq \hat{N}}.
\end{align}
Noticing the fact that $\theta_0(z)=1$ and performing a row expansion on the determinant (\ref{generaldeterminant}), we can deduce that
\begin{align}
	\begin{aligned}
		\sigma_n =&\frac{c_{\hat{N}}^{-1} c_{\delta_{\hat{N}}}^{-1}}{2^{\hat{N}(\hat{N}-1)}} A^{\frac{\delta_{\hat{N}}\left(\delta_{\hat{N}}+1\right)}{2}} B^{\frac{\hat{N}(\hat{N}+1)}{2}}\\ &\times\Theta_{\hat{N}}^{\left[l\right]}\left(\tilde{z}_1\right)\Theta_{\delta_{\hat{N}}}^{\left[\kappa\right]}\left(\tilde{z}_2\right)\left[1+O(A^{-2})\right].
	\end{aligned}
\end{align}
Therefore, we can conclude that $\sigma_1/\sigma_0\sim 1$, except at or near $(x,t)$ locations, where
\begin{align}
	\tilde{z}_{2,0}=A^{-1}\left(\tilde{x}_{2,0}-2\mathrm{i}\tilde{t}_{2,0}\right)	
\end{align}
is a nonzero root of the polynomial $\Theta_{\delta_N}^{[\kappa]}(\tilde{z}_2)$. The following proof of the middle region can be done through similar operations as Section \ref{sec5B}. Furthermore, the proofs for the outer region and the inner region can be carried out in a similar way.
\section{Conclusions}\label{sec6}
In summary, we explore some novel rogue waves patterns in the nonlocal NLS equation. Specifically, we show that when multiple free parameters get considerably large, its rogue wave patterns can approximately be predicted by the root structures of the Adler-Moser polynomials utilizing the aysmptotic analysis approach. Adler-Moser polynomials as a generalization of Yablonskii-Vorob'ev polynomial hierarchies have more free parameters, hence they have more diverse root structures, such as heart-shaped, fan-shaped and many others. In particular, we have divided the $(x,t)$-plane into three regions. In the outer and middle regions, the rogue wave patterns exhibit different distributions characterized by nonzero roots of two different Adler-Moser polynomials, while the inner region may contain a possible single rogue wave $u_{1010}(x, t),u_{1001}(x, t)$ or $u_{0110}(x, t)$. Many novel rogue wave patterns can be probed through the combination of patterns in different regions. We also provide some comparisons between the predicted patterns and the real ones, and numerically demonstrate that these predictions are in a good match with the actual results. At the moment we are submitting this paper, another type soliton, called \textit{Rogue Peakon} attracts our attention \cite{roguepeakon1}. Any nonlocal model has such peakons? We do not know yet, but we will keep on an eye on this new trend.
\section*{Acknowledgments}
The work was supported in part by the National Natural Science Foundation of China (No. 12171098, 11571079, 11701322), The Natural Science Foundation of Shanghai (No. 14ZR1403500), Shanghai Pujiang Program (No. 14PJD007).
\bibliographystyle{apsrev4-2}
\bibliography{references}

\begin{thebibliography}{48}%
\makeatletter
\providecommand \@ifxundefined [1]{%
 \@ifx{#1\undefined}
}%
\providecommand \@ifnum [1]{%
 \ifnum #1\expandafter \@firstoftwo
 \else \expandafter \@secondoftwo
 \fi
}%
\providecommand \@ifx [1]{%
 \ifx #1\expandafter \@firstoftwo
 \else \expandafter \@secondoftwo
 \fi
}%
\providecommand \natexlab [1]{#1}%
\providecommand \enquote  [1]{``#1''}%
\providecommand \bibnamefont  [1]{#1}%
\providecommand \bibfnamefont [1]{#1}%
\providecommand \citenamefont [1]{#1}%
\providecommand \href@noop [0]{\@secondoftwo}%
\providecommand \href [0]{\begingroup \@sanitize@url \@href}%
\providecommand \@href[1]{\@@startlink{#1}\@@href}%
\providecommand \@@href[1]{\endgroup#1\@@endlink}%
\providecommand \@sanitize@url [0]{\catcode `\\12\catcode `\$12\catcode
  `\&12\catcode `\#12\catcode `\^12\catcode `\_12\catcode `\%12\relax}%
\providecommand \@@startlink[1]{}%
\providecommand \@@endlink[0]{}%
\providecommand \url  [0]{\begingroup\@sanitize@url \@url }%
\providecommand \@url [1]{\endgroup\@href {#1}{\urlprefix }}%
\providecommand \urlprefix  [0]{URL }%
\providecommand \Eprint [0]{\href }%
\providecommand \doibase [0]{https://doi.org/}%
\providecommand \selectlanguage [0]{\@gobble}%
\providecommand \bibinfo  [0]{\@secondoftwo}%
\providecommand \bibfield  [0]{\@secondoftwo}%
\providecommand \translation [1]{[#1]}%
\providecommand \BibitemOpen [0]{}%
\providecommand \bibitemStop [0]{}%
\providecommand \bibitemNoStop [0]{.\EOS\space}%
\providecommand \EOS [0]{\spacefactor3000\relax}%
\providecommand \BibitemShut  [1]{\csname bibitem#1\endcsname}%
\let\auto@bib@innerbib\@empty
\bibitem [{\citenamefont {Dysthe}\ \emph {et~al.}(2008)\citenamefont {Dysthe},
  \citenamefont {Krogstad},\ and\ \citenamefont {M{\"u}ller}}]{khp}%
  \BibitemOpen
  \bibfield  {author} {\bibinfo {author} {\bibfnamefont {K.}~\bibnamefont
  {Dysthe}}, \bibinfo {author} {\bibfnamefont {H.~E.}\ \bibnamefont
  {Krogstad}},\ and\ \bibinfo {author} {\bibfnamefont {P.}~\bibnamefont
  {M{\"u}ller}},\ }\href@noop {} {\bibfield  {journal} {\bibinfo  {journal}
  {Annu. Rev. Fluid Mech.}\ }\textbf {\bibinfo {volume} {40}},\ \bibinfo
  {pages} {287} (\bibinfo {year} {2008})}\BibitemShut {NoStop}%
\bibitem [{\citenamefont {Bludov}\ \emph {et~al.}(2009)\citenamefont {Bludov},
  \citenamefont {Konotop},\ and\ \citenamefont {Akhmediev}}]{r8}%
  \BibitemOpen
  \bibfield  {author} {\bibinfo {author} {\bibfnamefont {Y.~V.}\ \bibnamefont
  {Bludov}}, \bibinfo {author} {\bibfnamefont {V.}~\bibnamefont {Konotop}},\
  and\ \bibinfo {author} {\bibfnamefont {N.}~\bibnamefont {Akhmediev}},\
  }\href@noop {} {\bibfield  {journal} {\bibinfo  {journal} {Phys. Rev. A}\
  }\textbf {\bibinfo {volume} {80}},\ \bibinfo {pages} {033610} (\bibinfo
  {year} {2009})}\BibitemShut {NoStop}%
\bibitem [{\citenamefont {Kibler}\ \emph {et~al.}(2010)\citenamefont {Kibler},
  \citenamefont {Fatome}, \citenamefont {Finot}, \citenamefont {Millot},
  \citenamefont {Dias}, \citenamefont {Genty}, \citenamefont {Akhmediev},\ and\
  \citenamefont {Dudley}}]{ki}%
  \BibitemOpen
  \bibfield  {author} {\bibinfo {author} {\bibfnamefont {B.}~\bibnamefont
  {Kibler}}, \bibinfo {author} {\bibfnamefont {J.}~\bibnamefont {Fatome}},
  \bibinfo {author} {\bibfnamefont {C.}~\bibnamefont {Finot}}, \bibinfo
  {author} {\bibfnamefont {G.}~\bibnamefont {Millot}}, \bibinfo {author}
  {\bibfnamefont {F.}~\bibnamefont {Dias}}, \bibinfo {author} {\bibfnamefont
  {G.}~\bibnamefont {Genty}}, \bibinfo {author} {\bibfnamefont
  {N.}~\bibnamefont {Akhmediev}},\ and\ \bibinfo {author} {\bibfnamefont
  {J.~M.}\ \bibnamefont {Dudley}},\ }\href@noop {} {\bibfield  {journal}
  {\bibinfo  {journal} {Nat. Phys.}\ }\textbf {\bibinfo {volume} {6}},\
  \bibinfo {pages} {790} (\bibinfo {year} {2010})}\BibitemShut {NoStop}%
\bibitem [{\citenamefont {Solli}\ \emph {et~al.}(2007)\citenamefont {Solli},
  \citenamefont {Ropers}, \citenamefont {Koonath},\ and\ \citenamefont
  {Jalali}}]{sol}%
  \BibitemOpen
  \bibfield  {author} {\bibinfo {author} {\bibfnamefont {D.~R.}\ \bibnamefont
  {Solli}}, \bibinfo {author} {\bibfnamefont {C.}~\bibnamefont {Ropers}},
  \bibinfo {author} {\bibfnamefont {P.}~\bibnamefont {Koonath}},\ and\ \bibinfo
  {author} {\bibfnamefont {B.}~\bibnamefont {Jalali}},\ }\href@noop {}
  {\bibfield  {journal} {\bibinfo  {journal} {Nature}\ }\textbf {\bibinfo
  {volume} {450}},\ \bibinfo {pages} {1054} (\bibinfo {year}
  {2007})}\BibitemShut {NoStop}%
\bibitem [{\citenamefont {Chabchoub}\ \emph {et~al.}(2011)\citenamefont
  {Chabchoub}, \citenamefont {Hoffmann},\ and\ \citenamefont
  {Akhmediev}}]{wb1}%
  \BibitemOpen
  \bibfield  {author} {\bibinfo {author} {\bibfnamefont {A.}~\bibnamefont
  {Chabchoub}}, \bibinfo {author} {\bibfnamefont {N.}~\bibnamefont
  {Hoffmann}},\ and\ \bibinfo {author} {\bibfnamefont {N.}~\bibnamefont
  {Akhmediev}},\ }\href@noop {} {\bibfield  {journal} {\bibinfo  {journal}
  {Phys. Rev. Lett.}\ }\textbf {\bibinfo {volume} {106}},\ \bibinfo {pages}
  {204502} (\bibinfo {year} {2011})}\BibitemShut {NoStop}%
\bibitem [{\citenamefont {Chabchoub}\ \emph
  {et~al.}(2012{\natexlab{a}})\citenamefont {Chabchoub}, \citenamefont
  {Hoffmann}, \citenamefont {Onorato},\ and\ \citenamefont {Akhmediev}}]{wb2}%
  \BibitemOpen
  \bibfield  {author} {\bibinfo {author} {\bibfnamefont {A.}~\bibnamefont
  {Chabchoub}}, \bibinfo {author} {\bibfnamefont {N.}~\bibnamefont {Hoffmann}},
  \bibinfo {author} {\bibfnamefont {M.}~\bibnamefont {Onorato}},\ and\ \bibinfo
  {author} {\bibfnamefont {N.}~\bibnamefont {Akhmediev}},\ }\href@noop {}
  {\bibfield  {journal} {\bibinfo  {journal} {Phys. Rev. X}\ }\textbf {\bibinfo
  {volume} {2}},\ \bibinfo {pages} {011015} (\bibinfo {year}
  {2012}{\natexlab{a}})}\BibitemShut {NoStop}%
\bibitem [{\citenamefont {Chabchoub}\ \emph
  {et~al.}(2012{\natexlab{b}})\citenamefont {Chabchoub}, \citenamefont
  {Hoffmann}, \citenamefont {Onorato}, \citenamefont {Slunyaev}, \citenamefont
  {Sergeeva}, \citenamefont {Pelinovsky},\ and\ \citenamefont
  {Akhmediev}}]{wb3}%
  \BibitemOpen
  \bibfield  {author} {\bibinfo {author} {\bibfnamefont {A.}~\bibnamefont
  {Chabchoub}}, \bibinfo {author} {\bibfnamefont {N.}~\bibnamefont {Hoffmann}},
  \bibinfo {author} {\bibfnamefont {M.}~\bibnamefont {Onorato}}, \bibinfo
  {author} {\bibfnamefont {A.}~\bibnamefont {Slunyaev}}, \bibinfo {author}
  {\bibfnamefont {A.}~\bibnamefont {Sergeeva}}, \bibinfo {author}
  {\bibfnamefont {E.}~\bibnamefont {Pelinovsky}},\ and\ \bibinfo {author}
  {\bibfnamefont {N.}~\bibnamefont {Akhmediev}},\ }\href@noop {} {\bibfield
  {journal} {\bibinfo  {journal} {Phys. Rev. E}\ }\textbf {\bibinfo {volume}
  {86}},\ \bibinfo {pages} {056601} (\bibinfo {year}
  {2012}{\natexlab{b}})}\BibitemShut {NoStop}%
\bibitem [{\citenamefont {Ganshin}\ \emph {et~al.}(2008)\citenamefont
  {Ganshin}, \citenamefont {Efimov}, \citenamefont {Kolmakov}, \citenamefont
  {Mezhov-Deglin},\ and\ \citenamefont {McClintock}}]{sf1}%
  \BibitemOpen
  \bibfield  {author} {\bibinfo {author} {\bibfnamefont {A.~N.}\ \bibnamefont
  {Ganshin}}, \bibinfo {author} {\bibfnamefont {V.~B.}\ \bibnamefont {Efimov}},
  \bibinfo {author} {\bibfnamefont {G.~V.}\ \bibnamefont {Kolmakov}}, \bibinfo
  {author} {\bibfnamefont {L.~P.}\ \bibnamefont {Mezhov-Deglin}},\ and\
  \bibinfo {author} {\bibfnamefont {P.~V.~E.}\ \bibnamefont {McClintock}},\
  }\href {https://doi.org/10.1103/PhysRevLett.101.065303} {\bibfield  {journal}
  {\bibinfo  {journal} {Phys. Rev. Lett.}\ }\textbf {\bibinfo {volume} {101}},\
  \bibinfo {pages} {065303} (\bibinfo {year} {2008})}\BibitemShut {NoStop}%
\bibitem [{\citenamefont {Efimov}\ \emph {et~al.}(2010)\citenamefont {Efimov},
  \citenamefont {Ganshin}, \citenamefont {Kolmakov}, \citenamefont
  {Mcclintock},\ and\ \citenamefont {Mezhov-Deglin}}]{sf2}%
  \BibitemOpen
  \bibfield  {author} {\bibinfo {author} {\bibfnamefont {V.~B.}\ \bibnamefont
  {Efimov}}, \bibinfo {author} {\bibfnamefont {A.~N.}\ \bibnamefont {Ganshin}},
  \bibinfo {author} {\bibfnamefont {G.~V.}\ \bibnamefont {Kolmakov}}, \bibinfo
  {author} {\bibfnamefont {P.~V.~E.}\ \bibnamefont {Mcclintock}},\ and\
  \bibinfo {author} {\bibfnamefont {L.~P.}\ \bibnamefont {Mezhov-Deglin}},\
  }\href@noop {} {\bibfield  {journal} {\bibinfo  {journal} {Eur. Phys. J.
  Spec. Top.}\ }\textbf {\bibinfo {volume} {185}},\ \bibinfo {pages} {181}
  (\bibinfo {year} {2010})}\BibitemShut {NoStop}%
\bibitem [{\citenamefont {Bailung}\ \emph {et~al.}(2011)\citenamefont
  {Bailung}, \citenamefont {Sharma},\ and\ \citenamefont {Nakamura}}]{pla}%
  \BibitemOpen
  \bibfield  {author} {\bibinfo {author} {\bibfnamefont {H.}~\bibnamefont
  {Bailung}}, \bibinfo {author} {\bibfnamefont {S.}~\bibnamefont {Sharma}},\
  and\ \bibinfo {author} {\bibfnamefont {Y.}~\bibnamefont {Nakamura}},\
  }\href@noop {} {\bibfield  {journal} {\bibinfo  {journal} {Phys. Rev. Lett.}\
  }\textbf {\bibinfo {volume} {107}},\ \bibinfo {pages} {255005} (\bibinfo
  {year} {2011})}\BibitemShut {NoStop}%
\bibitem [{\citenamefont {Akhmediev}\ \emph
  {et~al.}(2009{\natexlab{a}})\citenamefont {Akhmediev}, \citenamefont
  {Soto-Crespo},\ and\ \citenamefont {Ankiewicz}}]{akh}%
  \BibitemOpen
  \bibfield  {author} {\bibinfo {author} {\bibfnamefont {N.}~\bibnamefont
  {Akhmediev}}, \bibinfo {author} {\bibfnamefont {J.~M.}\ \bibnamefont
  {Soto-Crespo}},\ and\ \bibinfo {author} {\bibfnamefont {A.}~\bibnamefont
  {Ankiewicz}},\ }\href@noop {} {\bibfield  {journal} {\bibinfo  {journal}
  {Phys. Lett. A}\ }\textbf {\bibinfo {volume} {373}},\ \bibinfo {pages} {2137}
  (\bibinfo {year} {2009}{\natexlab{a}})}\BibitemShut {NoStop}%
\bibitem [{\citenamefont {Grimshaw}\ and\ \citenamefont {Tovbis}(2013)}]{gr}%
  \BibitemOpen
  \bibfield  {author} {\bibinfo {author} {\bibfnamefont {R.~H.~J.}\
  \bibnamefont {Grimshaw}}\ and\ \bibinfo {author} {\bibfnamefont
  {A.}~\bibnamefont {Tovbis}},\ }\href {https://doi.org/10.1098/rspa.2013.0094}
  {\bibfield  {journal} {\bibinfo  {journal} {Proc. R. Soc. A}\ }\textbf
  {\bibinfo {volume} {469}},\ \bibinfo {pages} {20130094} (\bibinfo {year}
  {2013})}\BibitemShut {NoStop}%
\bibitem [{\citenamefont {Peregrine}(1983)}]{pe}%
  \BibitemOpen
  \bibfield  {author} {\bibinfo {author} {\bibfnamefont {D.~H.}\ \bibnamefont
  {Peregrine}},\ }\href@noop {} {\bibfield  {journal} {\bibinfo  {journal} {J.
  Aust. Math. Soc. B}\ }\textbf {\bibinfo {volume} {25}},\ \bibinfo {pages}
  {16} (\bibinfo {year} {1983})}\BibitemShut {NoStop}%
\bibitem [{\citenamefont {Akhmediev}\ \emph
  {et~al.}(2009{\natexlab{b}})\citenamefont {Akhmediev}, \citenamefont
  {Ankiewicz},\ and\ \citenamefont {Soto-Crespo}}]{aa1}%
  \BibitemOpen
  \bibfield  {author} {\bibinfo {author} {\bibfnamefont {N.}~\bibnamefont
  {Akhmediev}}, \bibinfo {author} {\bibfnamefont {A.}~\bibnamefont
  {Ankiewicz}},\ and\ \bibinfo {author} {\bibfnamefont {J.~M.}\ \bibnamefont
  {Soto-Crespo}},\ }\href@noop {} {\bibfield  {journal} {\bibinfo  {journal}
  {Phys. Rev. E}\ }\textbf {\bibinfo {volume} {80}},\ \bibinfo {pages} {026601}
  (\bibinfo {year} {2009}{\natexlab{b}})}\BibitemShut {NoStop}%
\bibitem [{\citenamefont {Guo}\ \emph {et~al.}(2012)\citenamefont {Guo},
  \citenamefont {Ling},\ and\ \citenamefont {Liu}}]{g1}%
  \BibitemOpen
  \bibfield  {author} {\bibinfo {author} {\bibfnamefont {B.}~\bibnamefont
  {Guo}}, \bibinfo {author} {\bibfnamefont {L.}~\bibnamefont {Ling}},\ and\
  \bibinfo {author} {\bibfnamefont {Q.~P.}\ \bibnamefont {Liu}},\ }\href@noop
  {} {\bibfield  {journal} {\bibinfo  {journal} {Phys. Rev. E}\ }\textbf
  {\bibinfo {volume} {85}},\ \bibinfo {pages} {026607} (\bibinfo {year}
  {2012})}\BibitemShut {NoStop}%
\bibitem [{\citenamefont {Ohta}\ and\ \citenamefont
  {Yang}(2012{\natexlab{a}})}]{oh}%
  \BibitemOpen
  \bibfield  {author} {\bibinfo {author} {\bibfnamefont {Y.}~\bibnamefont
  {Ohta}}\ and\ \bibinfo {author} {\bibfnamefont {J.}~\bibnamefont {Yang}},\
  }\href@noop {} {\bibfield  {journal} {\bibinfo  {journal} {Proc. R. Soc. A}\
  }\textbf {\bibinfo {volume} {468}},\ \bibinfo {pages} {1716} (\bibinfo {year}
  {2012}{\natexlab{a}})}\BibitemShut {NoStop}%
\bibitem [{\citenamefont {Dubard}\ \emph {et~al.}(2010)\citenamefont {Dubard},
  \citenamefont {Gaillard}, \citenamefont {Klein},\ and\ \citenamefont
  {Matveev}}]{du}%
  \BibitemOpen
  \bibfield  {author} {\bibinfo {author} {\bibfnamefont {P.}~\bibnamefont
  {Dubard}}, \bibinfo {author} {\bibfnamefont {P.}~\bibnamefont {Gaillard}},
  \bibinfo {author} {\bibfnamefont {C.}~\bibnamefont {Klein}},\ and\ \bibinfo
  {author} {\bibfnamefont {V.}~\bibnamefont {Matveev}},\ }\href@noop {}
  {\bibfield  {journal} {\bibinfo  {journal} {Eur. Phys. J. Spec. Top.}\
  }\textbf {\bibinfo {volume} {185}},\ \bibinfo {pages} {247} (\bibinfo {year}
  {2010})}\BibitemShut {NoStop}%
\bibitem [{\citenamefont {Mu}\ and\ \citenamefont
  {Qin}(2014{\natexlab{a}})}]{mu2014}%
  \BibitemOpen
  \bibfield  {author} {\bibinfo {author} {\bibfnamefont {G.}~\bibnamefont
  {Mu}}\ and\ \bibinfo {author} {\bibfnamefont {Z.}~\bibnamefont {Qin}},\
  }\href@noop {} {\bibfield  {journal} {\bibinfo  {journal} {J. Phys. Soc.
  Jpn.}\ }\textbf {\bibinfo {volume} {83}},\ \bibinfo {pages} {104001}
  (\bibinfo {year} {2014}{\natexlab{a}})}\BibitemShut {NoStop}%
\bibitem [{\citenamefont {Ling}\ \emph {et~al.}(2014)\citenamefont {Ling},
  \citenamefont {Guo},\ and\ \citenamefont {Zhao}}]{manakov}%
  \BibitemOpen
  \bibfield  {author} {\bibinfo {author} {\bibfnamefont {L.}~\bibnamefont
  {Ling}}, \bibinfo {author} {\bibfnamefont {B.}~\bibnamefont {Guo}},\ and\
  \bibinfo {author} {\bibfnamefont {L.}~\bibnamefont {Zhao}},\ }\href@noop {}
  {\bibfield  {journal} {\bibinfo  {journal} {Phys. Rev. E}\ }\textbf {\bibinfo
  {volume} {89}},\ \bibinfo {pages} {041201} (\bibinfo {year}
  {2014})}\BibitemShut {NoStop}%
\bibitem [{\citenamefont {Ankiewicz}\ \emph {et~al.}(2010)\citenamefont
  {Ankiewicz}, \citenamefont {Soto-Crespo},\ and\ \citenamefont
  {Akhmediev}}]{aa2}%
  \BibitemOpen
  \bibfield  {author} {\bibinfo {author} {\bibfnamefont {A.}~\bibnamefont
  {Ankiewicz}}, \bibinfo {author} {\bibfnamefont {J.~M.}\ \bibnamefont
  {Soto-Crespo}},\ and\ \bibinfo {author} {\bibfnamefont {N.}~\bibnamefont
  {Akhmediev}},\ }\href {https://doi.org/10.1103/PhysRevE.81.046602} {\bibfield
   {journal} {\bibinfo  {journal} {Phys. Rev. E}\ }\textbf {\bibinfo {volume}
  {81}},\ \bibinfo {pages} {046602} (\bibinfo {year} {2010})}\BibitemShut
  {NoStop}%
\bibitem [{\citenamefont {Qin}\ and\ \citenamefont {Mu}(2012)}]{qin}%
  \BibitemOpen
  \bibfield  {author} {\bibinfo {author} {\bibfnamefont {Z.}~\bibnamefont
  {Qin}}\ and\ \bibinfo {author} {\bibfnamefont {G.}~\bibnamefont {Mu}},\
  }\href@noop {} {\bibfield  {journal} {\bibinfo  {journal} {Phys. Rev. E}\
  }\textbf {\bibinfo {volume} {86}},\ \bibinfo {pages} {036601} (\bibinfo
  {year} {2012})}\BibitemShut {NoStop}%
\bibitem [{\citenamefont {Guo}\ \emph {et~al.}(2013)\citenamefont {Guo},
  \citenamefont {Ling},\ and\ \citenamefont {Liu}}]{gll}%
  \BibitemOpen
  \bibfield  {author} {\bibinfo {author} {\bibfnamefont {B.}~\bibnamefont
  {Guo}}, \bibinfo {author} {\bibfnamefont {L.}~\bibnamefont {Ling}},\ and\
  \bibinfo {author} {\bibfnamefont {Q.}~\bibnamefont {Liu}},\ }\href@noop {}
  {\bibfield  {journal} {\bibinfo  {journal} {Stud. Appl. Math.}\ }\textbf
  {\bibinfo {volume} {130}},\ \bibinfo {pages} {317} (\bibinfo {year}
  {2013})}\BibitemShut {NoStop}%
\bibitem [{\citenamefont {Jimbo}\ and\ \citenamefont {Miwa}(1983)}]{mt}%
  \BibitemOpen
  \bibfield  {author} {\bibinfo {author} {\bibfnamefont {M.}~\bibnamefont
  {Jimbo}}\ and\ \bibinfo {author} {\bibfnamefont {T.}~\bibnamefont {Miwa}},\
  }\href@noop {} {\bibfield  {journal} {\bibinfo  {journal} {Publ. Res. Inst.
  Math. Sci.}\ }\textbf {\bibinfo {volume} {19}},\ \bibinfo {pages} {943}
  (\bibinfo {year} {1983})}\BibitemShut {NoStop}%
\bibitem [{\citenamefont {Hirota}(2004)}]{hirota}%
  \BibitemOpen
  \bibfield  {author} {\bibinfo {author} {\bibfnamefont {R.}~\bibnamefont
  {Hirota}},\ }\href@noop {} {\emph {\bibinfo {title} {The direct method in
  soliton theory}}},\ \bibinfo {number} {155}\ (\bibinfo  {publisher}
  {Cambridge university press},\ \bibinfo {year} {2004})\BibitemShut {NoStop}%
\bibitem [{\citenamefont {Ohta}\ and\ \citenamefont
  {Yang}(2012{\natexlab{b}})}]{yo1}%
  \BibitemOpen
  \bibfield  {author} {\bibinfo {author} {\bibfnamefont {Y.}~\bibnamefont
  {Ohta}}\ and\ \bibinfo {author} {\bibfnamefont {J.}~\bibnamefont {Yang}},\
  }\href@noop {} {\bibfield  {journal} {\bibinfo  {journal} {Phys. Rev. E}\
  }\textbf {\bibinfo {volume} {86}},\ \bibinfo {pages} {036604} (\bibinfo
  {year} {2012}{\natexlab{b}})}\BibitemShut {NoStop}%
\bibitem [{\citenamefont {Mu}\ and\ \citenamefont
  {Qin}(2014{\natexlab{b}})}]{mu2}%
  \BibitemOpen
  \bibfield  {author} {\bibinfo {author} {\bibfnamefont {G.}~\bibnamefont
  {Mu}}\ and\ \bibinfo {author} {\bibfnamefont {Z.}~\bibnamefont {Qin}},\
  }\href@noop {} {\bibfield  {journal} {\bibinfo  {journal} {Nonlinear Anal.
  Real World Appl.}\ }\textbf {\bibinfo {volume} {18}},\ \bibinfo {pages} {1}
  (\bibinfo {year} {2014}{\natexlab{b}})}\BibitemShut {NoStop}%
\bibitem [{\citenamefont {Wang}\ \emph {et~al.}(2023)\citenamefont {Wang},
  \citenamefont {Qin}, \citenamefont {Mu},\ and\ \citenamefont {Zheng}}]{wang}%
  \BibitemOpen
  \bibfield  {author} {\bibinfo {author} {\bibfnamefont {T.}~\bibnamefont
  {Wang}}, \bibinfo {author} {\bibfnamefont {Z.}~\bibnamefont {Qin}}, \bibinfo
  {author} {\bibfnamefont {G.}~\bibnamefont {Mu}},\ and\ \bibinfo {author}
  {\bibfnamefont {F.}~\bibnamefont {Zheng}},\ }\href@noop {} {\bibfield
  {journal} {\bibinfo  {journal} {Appl. Math. Lett.}\ }\textbf {\bibinfo
  {volume} {140}},\ \bibinfo {pages} {108571} (\bibinfo {year}
  {2023})}\BibitemShut {NoStop}%
\bibitem [{\citenamefont {Yang}\ \emph {et~al.}(2024)\citenamefont {Yang},
  \citenamefont {Mu},\ and\ \citenamefont {Qin}}]{kg}%
  \BibitemOpen
  \bibfield  {author} {\bibinfo {author} {\bibfnamefont {Z.}~\bibnamefont
  {Yang}}, \bibinfo {author} {\bibfnamefont {G.}~\bibnamefont {Mu}},\ and\
  \bibinfo {author} {\bibfnamefont {Z.}~\bibnamefont {Qin}},\ }\href@noop {}
  {\bibfield  {journal} {\bibinfo  {journal} {Chaos}\ }\textbf {\bibinfo
  {volume} {34}} (\bibinfo {year} {2024})}\BibitemShut {NoStop}%
\bibitem [{\citenamefont {Ablowitz}\ and\ \citenamefont {Segur}(1981)}]{ins}%
  \BibitemOpen
  \bibfield  {author} {\bibinfo {author} {\bibfnamefont {M.~J.}\ \bibnamefont
  {Ablowitz}}\ and\ \bibinfo {author} {\bibfnamefont {H.}~\bibnamefont
  {Segur}},\ }\href@noop {} {\emph {\bibinfo {title} {Solitons and the inverse
  scattering transform}}}\ (\bibinfo  {publisher} {SIAM},\ \bibinfo {year}
  {1981})\BibitemShut {NoStop}%
\bibitem [{\citenamefont {Kedziora}\ \emph {et~al.}(2011)\citenamefont
  {Kedziora}, \citenamefont {Ankiewicz},\ and\ \citenamefont
  {Akhmediev}}]{NLSPa1}%
  \BibitemOpen
  \bibfield  {author} {\bibinfo {author} {\bibfnamefont {D.~J.}\ \bibnamefont
  {Kedziora}}, \bibinfo {author} {\bibfnamefont {A.}~\bibnamefont
  {Ankiewicz}},\ and\ \bibinfo {author} {\bibfnamefont {N.}~\bibnamefont
  {Akhmediev}},\ }\href@noop {} {\bibfield  {journal} {\bibinfo  {journal}
  {Phys. Rev. E}\ }\textbf {\bibinfo {volume} {84}},\ \bibinfo {pages} {056611}
  (\bibinfo {year} {2011})}\BibitemShut {NoStop}%
\bibitem [{\citenamefont {Kedziora}\ \emph {et~al.}(2013)\citenamefont
  {Kedziora}, \citenamefont {Ankiewicz},\ and\ \citenamefont
  {Akhmediev}}]{NLSPa2}%
  \BibitemOpen
  \bibfield  {author} {\bibinfo {author} {\bibfnamefont {D.~J.}\ \bibnamefont
  {Kedziora}}, \bibinfo {author} {\bibfnamefont {A.}~\bibnamefont
  {Ankiewicz}},\ and\ \bibinfo {author} {\bibfnamefont {N.}~\bibnamefont
  {Akhmediev}},\ }\href@noop {} {\bibfield  {journal} {\bibinfo  {journal}
  {Phys. Rev. E}\ }\textbf {\bibinfo {volume} {88}},\ \bibinfo {pages} {013207}
  (\bibinfo {year} {2013})}\BibitemShut {NoStop}%
\bibitem [{\citenamefont {Yang}\ and\ \citenamefont
  {Yang}(2021{\natexlab{a}})}]{NLSYV}%
  \BibitemOpen
  \bibfield  {author} {\bibinfo {author} {\bibfnamefont {B.}~\bibnamefont
  {Yang}}\ and\ \bibinfo {author} {\bibfnamefont {J.}~\bibnamefont {Yang}},\
  }\href@noop {} {\bibfield  {journal} {\bibinfo  {journal} {Phys. D}\ }\textbf
  {\bibinfo {volume} {419}},\ \bibinfo {pages} {132850} (\bibinfo {year}
  {2021}{\natexlab{a}})}\BibitemShut {NoStop}%
\bibitem [{\citenamefont {Yang}\ and\ \citenamefont
  {Yang}(2021{\natexlab{b}})}]{uniYV}%
  \BibitemOpen
  \bibfield  {author} {\bibinfo {author} {\bibfnamefont {B.}~\bibnamefont
  {Yang}}\ and\ \bibinfo {author} {\bibfnamefont {J.}~\bibnamefont {Yang}},\
  }\href@noop {} {\bibfield  {journal} {\bibinfo  {journal} {Phys. D}\ }\textbf
  {\bibinfo {volume} {425}},\ \bibinfo {pages} {132958} (\bibinfo {year}
  {2021}{\natexlab{b}})}\BibitemShut {NoStop}%
\bibitem [{\citenamefont {Ling}\ and\ \citenamefont {Su}(2024)}]{OM2}%
  \BibitemOpen
  \bibfield  {author} {\bibinfo {author} {\bibfnamefont {L.}~\bibnamefont
  {Ling}}\ and\ \bibinfo {author} {\bibfnamefont {H.}~\bibnamefont {Su}},\
  }\href@noop {} {\bibfield  {journal} {\bibinfo  {journal} {Phys. D}\ }\textbf
  {\bibinfo {volume} {461}},\ \bibinfo {pages} {134111} (\bibinfo {year}
  {2024})}\BibitemShut {NoStop}%
\bibitem [{\citenamefont {Lin}\ and\ \citenamefont {Ling}(2024)}]{gWH2}%
  \BibitemOpen
  \bibfield  {author} {\bibinfo {author} {\bibfnamefont {H.}~\bibnamefont
  {Lin}}\ and\ \bibinfo {author} {\bibfnamefont {L.}~\bibnamefont {Ling}},\
  }\href@noop {} {\bibfield  {journal} {\bibinfo  {journal} {Chaos}\ }\textbf
  {\bibinfo {volume} {34}} (\bibinfo {year} {2024})}\BibitemShut {NoStop}%
\bibitem [{\citenamefont {Yang}\ and\ \citenamefont {Yang}(2024)}]{AM1}%
  \BibitemOpen
  \bibfield  {author} {\bibinfo {author} {\bibfnamefont {B.}~\bibnamefont
  {Yang}}\ and\ \bibinfo {author} {\bibfnamefont {J.}~\bibnamefont {Yang}},\
  }\href@noop {} {\bibfield  {journal} {\bibinfo  {journal} {Appl. Math.
  Lett.}\ }\textbf {\bibinfo {volume} {148}},\ \bibinfo {pages} {108871}
  (\bibinfo {year} {2024})}\BibitemShut {NoStop}%
\bibitem [{\citenamefont {Zhang}\ and\ \citenamefont {Wu}(2024)}]{NLNLSYV}%
  \BibitemOpen
  \bibfield  {author} {\bibinfo {author} {\bibfnamefont {G.}~\bibnamefont
  {Zhang}}\ and\ \bibinfo {author} {\bibfnamefont {C.}~\bibnamefont {Wu}},\
  }\href@noop {} {\bibfield  {journal} {\bibinfo  {journal} {Phys. Fluids}\
  }\textbf {\bibinfo {volume} {36}} (\bibinfo {year} {2024})}\BibitemShut
  {NoStop}%
\bibitem [{\citenamefont {Ablowitz}\ and\ \citenamefont
  {Musslimani}(2013)}]{NLNLS1}%
  \BibitemOpen
  \bibfield  {author} {\bibinfo {author} {\bibfnamefont {M.~J.}\ \bibnamefont
  {Ablowitz}}\ and\ \bibinfo {author} {\bibfnamefont {Z.~H.}\ \bibnamefont
  {Musslimani}},\ }\href@noop {} {\bibfield  {journal} {\bibinfo  {journal}
  {Phys. Rev. Lett.}\ }\textbf {\bibinfo {volume} {110}},\ \bibinfo {pages}
  {064105} (\bibinfo {year} {2013})}\BibitemShut {NoStop}%
\bibitem [{\citenamefont {Ablowitz}\ and\ \citenamefont
  {Musslimani}(2016)}]{NLNLS2}%
  \BibitemOpen
  \bibfield  {author} {\bibinfo {author} {\bibfnamefont {M.~J.}\ \bibnamefont
  {Ablowitz}}\ and\ \bibinfo {author} {\bibfnamefont {Z.~H.}\ \bibnamefont
  {Musslimani}},\ }\href@noop {} {\bibfield  {journal} {\bibinfo  {journal}
  {Nonlinearity}\ }\textbf {\bibinfo {volume} {29}},\ \bibinfo {pages} {915}
  (\bibinfo {year} {2016})}\BibitemShut {NoStop}%
\bibitem [{\citenamefont {Wen}\ \emph {et~al.}(2016)\citenamefont {Wen},
  \citenamefont {Yan},\ and\ \citenamefont {Yang}}]{NLNLS3}%
  \BibitemOpen
  \bibfield  {author} {\bibinfo {author} {\bibfnamefont {X.~Y.}\ \bibnamefont
  {Wen}}, \bibinfo {author} {\bibfnamefont {Z.}~\bibnamefont {Yan}},\ and\
  \bibinfo {author} {\bibfnamefont {Y.}~\bibnamefont {Yang}},\ }\href
  {https://doi.org/10.1063/1.4954767} {\bibfield  {journal} {\bibinfo
  {journal} {Chaos}\ }\textbf {\bibinfo {volume} {26}},\ \bibinfo {pages}
  {063123} (\bibinfo {year} {2016})}\BibitemShut {NoStop}%
\bibitem [{\citenamefont {Gerdjikov}\ and\ \citenamefont
  {Saxena}(2017)}]{NLNLS5}%
  \BibitemOpen
  \bibfield  {author} {\bibinfo {author} {\bibfnamefont {V.~S.}\ \bibnamefont
  {Gerdjikov}}\ and\ \bibinfo {author} {\bibfnamefont {A.}~\bibnamefont
  {Saxena}},\ }\href {https://doi.org/10.1063/1.4974018} {\bibfield  {journal}
  {\bibinfo  {journal} {J. Math. Phys.}\ }\textbf {\bibinfo {volume} {58}},\
  \bibinfo {pages} {013502} (\bibinfo {year} {2017})}\BibitemShut {NoStop}%
\bibitem [{\citenamefont {Konotop}\ \emph {et~al.}(2016)\citenamefont
  {Konotop}, \citenamefont {Yang},\ and\ \citenamefont {Zezyulin}}]{PTsym}%
  \BibitemOpen
  \bibfield  {author} {\bibinfo {author} {\bibfnamefont {V.~V.}\ \bibnamefont
  {Konotop}}, \bibinfo {author} {\bibfnamefont {J.}~\bibnamefont {Yang}},\ and\
  \bibinfo {author} {\bibfnamefont {D.~A.}\ \bibnamefont {Zezyulin}},\
  }\href@noop {} {\bibfield  {journal} {\bibinfo  {journal} {Rev. Mod. Phys.}\
  }\textbf {\bibinfo {volume} {88}},\ \bibinfo {pages} {035002} (\bibinfo
  {year} {2016})}\BibitemShut {NoStop}%
\bibitem [{\citenamefont {Yang}\ and\ \citenamefont {Yang}(2019)}]{NLNLSDT}%
  \BibitemOpen
  \bibfield  {author} {\bibinfo {author} {\bibfnamefont {B.}~\bibnamefont
  {Yang}}\ and\ \bibinfo {author} {\bibfnamefont {J.}~\bibnamefont {Yang}},\
  }\href@noop {} {\bibfield  {journal} {\bibinfo  {journal} {Lett. Math.
  Phys.}\ }\textbf {\bibinfo {volume} {109}},\ \bibinfo {pages} {945} (\bibinfo
  {year} {2019})}\BibitemShut {NoStop}%
\bibitem [{\citenamefont {Yang}\ and\ \citenamefont {Yang}(2020)}]{NLNLSKP}%
  \BibitemOpen
  \bibfield  {author} {\bibinfo {author} {\bibfnamefont {B.}~\bibnamefont
  {Yang}}\ and\ \bibinfo {author} {\bibfnamefont {J.}~\bibnamefont {Yang}},\
  }\href@noop {} {\bibfield  {journal} {\bibinfo  {journal} {J. Math. Anal.
  Appl.}\ }\textbf {\bibinfo {volume} {487}},\ \bibinfo {pages} {124023}
  (\bibinfo {year} {2020})}\BibitemShut {NoStop}%
\bibitem [{\citenamefont {Adler}\ and\ \citenamefont {Moser}(1978)}]{AMpoly}%
  \BibitemOpen
  \bibfield  {author} {\bibinfo {author} {\bibfnamefont {M.}~\bibnamefont
  {Adler}}\ and\ \bibinfo {author} {\bibfnamefont {J.}~\bibnamefont {Moser}},\
  }\href@noop {} {\bibfield  {journal} {\bibinfo  {journal} {Comm. Math.
  Phys.}\ }\textbf {\bibinfo {volume} {61}},\ \bibinfo {pages} {1} (\bibinfo
  {year} {1978})}\BibitemShut {NoStop}%
\bibitem [{\citenamefont {Aref}(2007)}]{AMpoly2}%
  \BibitemOpen
  \bibfield  {author} {\bibinfo {author} {\bibfnamefont {H.}~\bibnamefont
  {Aref}},\ }\href {https://doi.org/10.1016/j.fluiddyn.2006.04.004} {\bibfield
  {journal} {\bibinfo  {journal} {Fluid Dyn. Res.}\ }\textbf {\bibinfo {volume}
  {39}},\ \bibinfo {pages} {5} (\bibinfo {year} {2007})}\BibitemShut {NoStop}%
\bibitem [{\citenamefont {Clarkson}(2009)}]{AMpoly3}%
  \BibitemOpen
  \bibfield  {author} {\bibinfo {author} {\bibfnamefont {P.~A.}\ \bibnamefont
  {Clarkson}},\ }\href@noop {} {\bibfield  {journal} {\bibinfo  {journal}
  {Stud. Appl. Math.}\ }\textbf {\bibinfo {volume} {123}},\ \bibinfo {pages}
  {37} (\bibinfo {year} {2009})}\BibitemShut {NoStop}%
\bibitem [{\citenamefont {Zhu}\ \emph {et~al.}(2023)\citenamefont {Zhu},
  \citenamefont {Zeng}, \citenamefont {Jiang}, \citenamefont {Xia},\ and\
  \citenamefont {Qiao}}]{roguepeakon1}%
  \BibitemOpen
  \bibfield  {author} {\bibinfo {author} {\bibfnamefont {M.}~\bibnamefont
  {Zhu}}, \bibinfo {author} {\bibfnamefont {Z.}~\bibnamefont {Zeng}}, \bibinfo
  {author} {\bibfnamefont {Z.}~\bibnamefont {Jiang}}, \bibinfo {author}
  {\bibfnamefont {B.}~\bibnamefont {Xia}},\ and\ \bibinfo {author}
  {\bibfnamefont {Z.}~\bibnamefont {Qiao}},\ }\href@noop {} {\bibfield
  {journal} {\bibinfo  {journal} {Arxiv}\ }\textbf {\bibinfo {volume}
  {abs/2308.11508}} (\bibinfo {year} {2023})},\ \Eprint
  {https://arxiv.org/abs/2308.11508} {arXiv:2308.11508 [nlin.SI]} \BibitemShut
  {NoStop}%
\end{thebibliography}%
\end{document}